\documentclass[journal=jctcce,layout=twocolumn,manuscript=article]{achemso}
\usepackage{amsmath}
\usepackage{graphicx}
\usepackage{color}

\makeatletter
\let\l@addto@macro\relax
\makeatother
\let\oldmaketitle\maketitle
\let\maketitle\relax

\author{Sabry G. Moustafa}
\email{smoustaf@trinity.edu}
\affiliation{Department of Engineering Science, Trinity University, San Antonio, Texas 78212, USA}
\author{Andrew J. Schultz}
\affiliation{Department of Chemical and Biological Engineering, University at Buffalo, The State University of New York, Buffalo, New York 14260-4200, USA}
\title{Harmonic Oscillator Staging Coordinates for Efficient Path Integral Simulations of Quantum Oscillators and Crystals}

\usepackage[fontsize=9pt]{scrextend}

\usepackage{xr}
\makeatletter

\newcommand*{\addFileDependency}[1]{
\typeout{(#1)}
\@addtofilelist{#1}
\IfFileExists{#1}{}{\typeout{No file #1.}}
}\makeatother
\newcommand*{\myexternaldocument}[1]{%
\externaldocument{#1}%
\addFileDependency{#1.tex}%
\addFileDependency{#1.aux}%
}
\myexternaldocument{si}

\usepackage{amsmath}

\begin{document}
\twocolumn[
\begin{@twocolumnfalse}
\oldmaketitle
Imaginary-time path integral (PI) is a rigorous quantum mechanical tool to compute static properties at finite temperatures. However, the stiff nature of the internal PI modes poses a sampling challenge. This is commonly tackled using staging coordinates, in which the free particle (FP) contribution of the PI action is diagonalized. We introduce novel and simple staging coordinates that diagonalize the entire action of the harmonic oscillator (HO) model, rendering it efficiently applicable to (\textit{exclusively}) systems with a harmonic character, such as quantum oscillators and crystals. The method is not applicable to fluids or systems with imaginary modes. Unlike FP staging, the HO staging provides a unique treatment of the centroid mode. We provide implementation schemes for PIMC and PIMD simulations in NVT ensemble. Sampling efficiency is assessed in terms of the precision and accuracy of estimating the energy and heat capacity of a one-dimensional HO and an asymmetric anharmonic oscillator (AO). In PIMC, the HO coordinates propose collective moves that perfectly sample the HO contribution, then (for AO) the residual anharmonic term is sampled using standard Metropolis method. This results in a high acceptance rate and, hence, high precision, in comparison to the FP staging. In PIMD, the HO coordinates naturally prescribe definitions for the fictitious masses, yielding equal frequencies of all modes when applied to the HO model. This allows for a substantially larger time step sizes relative to standard staging, without affecting accuracy nor integrator stability. For completeness, we also present results using normal mode (NM) coordinates, based on both HO and FP models. While staging and NM coordinates show similar performance (for FP or HO), staging is computationally preferable due to its cheaper scaling with the number of beads. The simplicity and the enhanced sampling gained by the HO coordinates open avenues for efficient estimation of nuclear quantum effects in more complex systems with a harmonic character, such as real molecular bonds and quantum crystals.\\
\end{@twocolumnfalse}]

\section{INTRODUCTION}
Nuclear quantum effects of distinguishable particles, such as zero-point energy and tunneling, play an important role when the thermal energy becomes smaller than the spacing between quantum energy levels. This is especially relevant to lightweight elements (e.g., hydrogen and helium) and/or high frequency (stiff) modes.\cite{ceriotti2018review} For crystalline systems, an often adopted quantitative threshold for these effects is the Debye temperature, which is proportional to the maximum phonon frequency present.\cite{herrero2014path} In these cases, quantum treatment for thermodynamic properties is crucial. According to quantum statistical mechanics, the canonical partition function of a system at a temperature $T$ is given as the trace of its density matrix, $Z = {\rm tr}\left(e^{-\beta {\hat H}}\right)$, where $\hat H$ is the Hamiltonian operator and $\beta$ is the reciprocal of the thermal energy, $k_BT$, with $k_{\rm B}$ the Boltzmann constant. However, except for simple cases, such as free particles and harmonic oscillators, it is challenging to determine $Z$ analytically. In this regard, Feynman's imaginary-time path integral (PI) formulation provides a numerically tractable alternative.\cite{Feynman1965} In this framework, the quantum system is mapped onto an extended classical isomorphism, made of $n$ replicas (imaginary time slices) of the original system. These images are connected consecutively through harmonic springs, in a closed ring-polymer arrangement of $n$ beads. The true partition function is then recovered in the $n \to \infty$ limit.

The ``standard'' path integral primitive action uses the free-particle (FP) density matrix as a reference and adopts a second-order Trotter factorization. In this approximation, the partition function of a system of $N$ distinguishable particles, occupying $d$-dimensional space, is given by the following integral over a $dNn$ extended phase space,\cite{tuckerman2023book}
\begin{subequations}
\label{eq:ZV_PI}
\begin{align}
\label{eq:Z_PI}
Z\left(\beta \right) &= \left(\frac{m n}{2\pi \hbar^2 \beta}\right)^{dNn/2} \int {\rm d} {\bf x} 
\; {\rm exp}\left(-\beta V\left({\bf x}, \beta\right)\right), \\
\label{eq:V_PI}
V \left({\bf x}, \beta\right) &=  \frac{1}{n}\sum_{i=0}^{n-1} \frac{1}{2} m \omega_n^2\left({\bf x}_i - {\bf x}_{i-1} \right)^2 + \frac{1}{n} \sum_{i=0}^{n-1}   U\left({\bf x}_i\right),
\end{align}    
\end{subequations}
where $m$ is the atom mass, $\hbar\equiv h/2\pi$ is the reduced Planck constant, $\omega_n \equiv n/\beta\hbar$, and $\bf x$ is a vector of all the $dNn$ beads coordinates, with ${\bf x}_{i}$ representing the coordinates pf the $i^{\rm th}$ replica (of length $dN$). Here, $V$ represents a $T$-dependent effective potential, which comprises the intermolecular potentials $U\left({\bf x}_i\right)$, from each replica, and the kinetic term, represented by FP harmonic interactions of identical spring constants, $m\omega_n^2/n$. 

In dense phases, $\beta\hbar\omega_{\rm max}$ is often used to gauge ``quantumness'' level, where $\omega_{\rm max}$ is the maximum  angular frequency in the system (Debye frequency in crystals).\cite{herrero2014path} On the other hand, numerical convergence in PI is achieved by setting the number of beads equal to multiples of the quantumness, such that $n = C \beta\hbar\omega_{\rm max}$, with $C$ being a constant greater than one.\cite{ceriotti2018review} Hence, based on the definition of $\omega_n$, the spring constant of the kinetic term, $m\omega_n^2/n$, is stiffer than the maximum intermolecular counterpart, $m\omega_{\rm max}^2/n$, hence, $\omega_n\gg\omega_{\rm max}$ at convergence. The relatively stiff harmonic interactions are known to cause sampling challenges in both PIMD and PIMC simulations, using Cartesian coordinates.\cite{tuckerman1993efficient} For PIMD, ergodicity problems arise due to the dominant harmonic behavior, which has a poor energy exchange between different modes.\cite{hall1984nonergodicity} Nevertheless, this is usually alleviated using specific thermostatting methods, such as Langevin and Nose-Hoover chain thermostats. More importantly, small time steps are necessary in order to resolve the fast stiff modes, which clearly introduces slow statistical convergence. For PIMC, on the other hand, the local nature of the MC moves (one bead at a time) and their small size (to handle stiff potentials) cause sampling inefficiency.\cite{tuckerman1993efficient} Since $\omega_n$ is proportional to $n$, then these issues are more relevant at low temperatures, where more beads are needed for convergence.

These challenges are primarily due to using Cartesian coordinates, which are inherently local and, hence, do not account for the collective character of the system modes. A natural solution is then to transform the coordinates to a new set compatible with the PI action. The standard approaches are either to use staging~\cite{tuckerman1996efficient,tuckerman1993efficient} or normal mode (NM)~\cite{herman1982path,cao1993born} coordinates, which diagonalize the kinetic contribution of the PI action (i.e., FP model). In application to PIMD, fictitious (or, dynamic) masses are assigned to the transformed coordinates, in an attempt to narrow the time scale differences and, hence, improve sampling. The system dynamics will clearly change; however, this does not affect static properties, such as energy and heat capacity. For PIMC, a collective move of all beads associated with a given atom is proposed according to independent Gaussian distributions, which describe the kinetic term in the transformed coordinates. Then, the move is either accepted or rejected according to the usual Metropolis scheme, applied to the residual (off kinetic) energy contribution. Although both coordinates diagonalize the same function, the appealing feature of staging is its ${\cal O}\left(n\right)$ computational scaling. This is in contrast to the NM case where the cost is ${\cal O}\left(n^2\right)$, or ${\cal O}\left(n\log\left(n\right)\right)$ using the fast Fourier transform (FFT) approach.

A common limitation of the coordinates lies in the fact that both coordinates describe only the internal degrees of freedom within the ring-polymer. However, traditional approaches are typically adopted to describe the external degree of freedom, which is represented by the centroid and a bead position for the case of NM and staging coordinates, respectively. For example, a typical practice in PIMD simulation is to assign the physical mass to that mode. On the other hand, for PIMC, the external degree of freedom is sampled separately using a translation move of the whole ring-polymer, according to the standard Metropolis scheme. Clearly, both of these choices are not system-dependent. However, taking into account the nature of the system being considered could, in principle, improve sampling efficiency.

We consider applications to systems with harmonic character, such as quantum oscillators. For crystals, the atomic vibrations could be approximated as independent oscillators, forming an Einstein crystal. In these cases, the harmonic contribution could be described using the known action of the harmonic oscillator (HO) model. In this study, we introduce a set of staging and NM coordinates, which diagonalize the entire HO action rather than merely the kinetic term, as adopted in the FP methods. Hence, we denote these methods as ``HO staging'' and ``HO NM'' coordinates. Unlike FP approaches, the new coordinates represent both internal and external degrees of freedom and, hence, provide a better description of the real system. In application to PIMD, the approach provides a natural prescription for the fictitious masses, including those of the centroid mode. Consequently, all modes fluctuate on the same time scale in the case of HO model, or nearly similar time scales with anharmonic systems. This, then, would allow for using relatively large time step sizes, without affecting the accuracy nor the MD integrator stability. This, in turn, results in higher precision estimates, for a given number of steps. For PIMC, the HO coordinates prescribes a collective move that samples perfectly ($100\%$ acceptance) the HO contribution of the PI action. Unlike FP moves, the centroid degree of freedom is inherently included with the new approach. The move is then accepted/rejected using the standard Metropolis criterion, according to the residual, anharmonic, contribution. This would then result in a high overall acceptance rate and, hence, higher precision in measured properties. 

To assess the performance, we apply the HO coordinates approach to measure the energy and heat capacity of a one-dimensional HO, in both PIMC and PIMD simulations, using the centroid virial estimator. Despite the model simplicity, it shares the same challenges encountered by PI simulations (stiff harmonic bonds, multiple time scales, nonergodicity, etc.).  In addition, the model has an exact solution, which is used as a means to check accuracy. However, the effect of anharmonicity was also investigated using an asymmetric anharmonic oscillator (AO), in which the HO is augmented by a quartic term. The statistical uncertainty in the measured PI estimators is used as a gauge to the sampling efficiency. In all cases, the HO coordinates show superior precision relative to existing FP coordinates. This is a manifestation of the method's high acceptance rate and large time step size allowed for PIMC and PIMD simulations, respectively. Similar to the FP method, both HO staging and NM coordinates show similar performance. We also explore the dependence of sampling efficiency on the number of beads, $n$. A typical behavior of the centroid virial estimator is observed, in which the statistical uncertainty shows no sensitivity to the number of beads after some value (significantly smaller than the $n=C\beta\hbar\omega_{\rm max}$ converged value). 

It is worth pointing out to another, less-common, second-order factorization, which uses the HO (rather than FP) propagator as a reference to formulate the discretized PI.\cite{friesner1984optimized,martyna2007low} The primary advantage of this method is its relatively small (or non for HO) finite-size effects compared to primitive PI (Eq.~\ref{eq:ZV_PI}). Nevertheless, since the ultimate interest is in the continuum limit ($n\to \infty$), extrapolation is required in both PI formulations. The primitive PI formulation, however, offers a few advantages over the HO-based PI method, prompting our choice to it. For example, in addition to its simplicity, the HO spring constant is independent of temperature, unlike the case with the HO-based PI.\cite{martyna2007low} This adds complexity to the analysis, especially when deriving properties based on temperature derivatives, such as heat capacity (considered here).  

The structure of this paper is as follows. In Sec.~\ref{sec:methods}, we first introduce the HO staging and normal mode coordinates formulation as a approach to diagonalize the HO action. This is followed by providing schemes for implementing these coordinates in PIMC and PIMD simulations. For PIMD, we define the new fictitious masses associated with the HO coordinates. We then provide formulas for the total energy and heat capacity, based on the centroid virial estimator, along with uncertainty analysis for both quantities. We follow this by detailed model and simulation parameters considered. In Sec.\ref{sec:results}, we apply the new coordinates to both HO and AO models and show performance results based on both PIMC and PIMD techniques. We then conclude in Sec.~\ref{sec:conclusion} with a summary the framework and its outcomes, along with potential future directions.

\section{FORMALISM AND METHODS}
\label{sec:methods}

\subsection{Harmonic Oscillator Staging}
\begin{figure}
\includegraphics[width=0.48\textwidth]{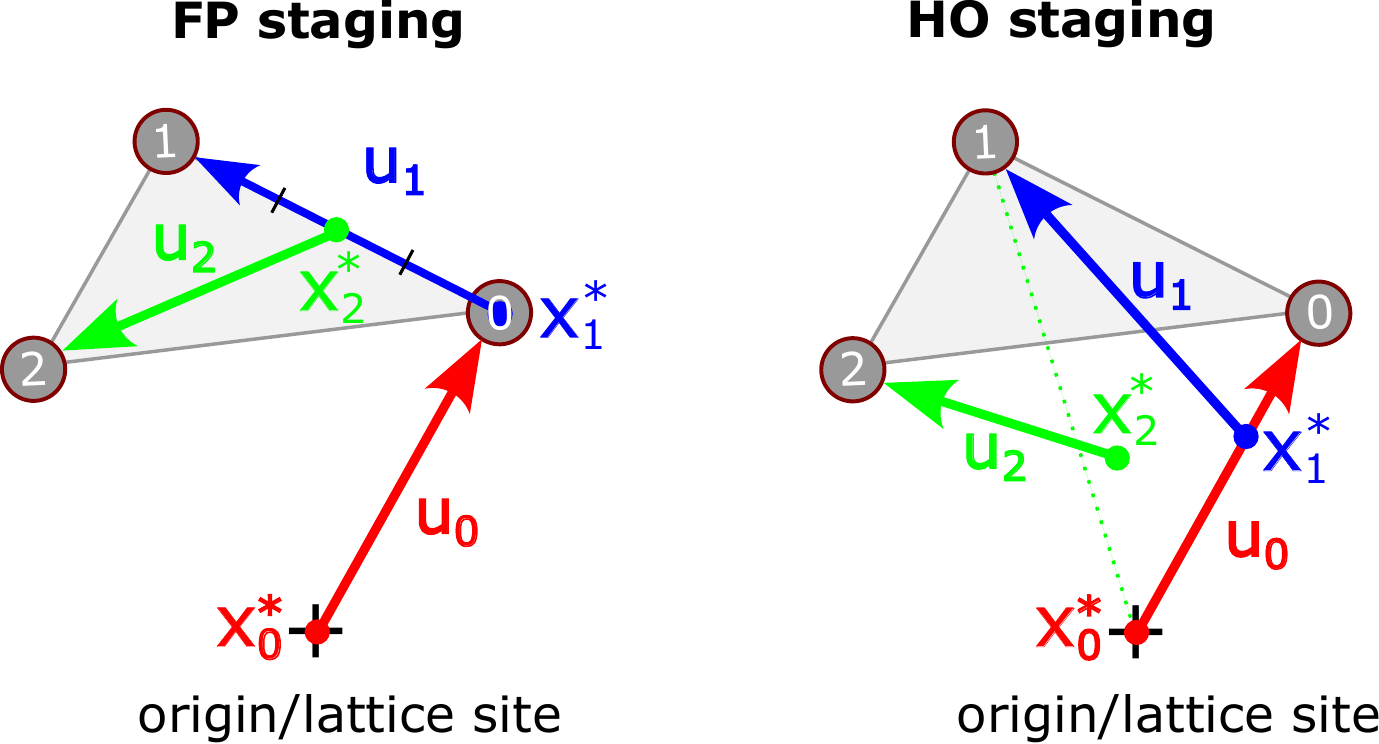}
\caption{A simplified example of the FP (left) and HO (right) staging coordinates $u_i$ (arrows) for a three-beads system. The instantaneous equilibrium position $x_0^*$ is always located at the global minimum (origin/lattice position). For $i>0$, $x_i^*$ for the FP staging lays on the line connecting beads $0$ and $i-1$ according to Eq.~\ref{eq:r2x_transformation_FP}, whereas, for HO staging it is contained within the triangle connecting the origin/lattice position and beads $0$ and $i-1$, according to   
by Eq.~\ref{eq:xc_ho_staging}. For example, $x_2^*$ is located within the origin-0-1 triangle, with the 0-1 side is connected by a dashed green line for clarity.}
\centering
\label{fig:staging_all}
\end{figure}
We will use the HO model to derive a new set of staging coordinates relevant to systems with a harmonic character, such as quantum oscillators and crystals. For simplicity of notation, we will develop the formulation for a one-dimensional HO. Extension to an arbitrary system size and/or dimensions is trivial due to the non-interacting nature of the oscillators, in all directions. According to Eq.~\ref{eq:V_PI}, the effective potential of the HO model is given by, 
\begin{eqnarray}
\label{eq:V_HO_real}
V\left({\bf x},\beta\right)  = \frac{1}{n} \sum_{i=0}^{n-1} \frac{1}{2} m \omega_n^2  \left(x_i -  x_{i-1} \right)^2
+ \frac{1}{n} \sum_{i=0}^{n-1} \frac{1}{2} m \omega^2   x_i^2,
\end{eqnarray}
where $x_i$ is the displacement of the $i^{\rm th}$ bead from its equilibrium position (lattice site for crystals), and $\omega$ is the HO angular frequency. The objective now is to diagonalize this coupled form within the staging framework. To accomplish this, we define the following set of generalized staging coordinates:
\begin{align}
\label{eq:ui_def}
u_i\left(x_0,x_1,\dots,x_i\right) &\equiv x_i - x_i^*\left(x_0,x_1,\dots,x_{i-1}\right), \; i>0
\end{align}
where $x_i^*$ is an instantaneous ``equilibrium''  position, solely dependent on the positions of the previous beads ($0,\dots,i-1$). We define the zero staging coordinate as $u_0=x_0$ (hence, $x_0^*=0$), which acts as an external mode, somewhat similar to the centroid. We then use these coordinates to express the effective PI potential (Eq.~\ref{eq:V_HO_real}) in the following decoupled form of independent harmonic oscillators:
\begin{align}
\label{eq:V_HO_staging}
    V\left({\bf u},\beta\right) &= \sum_{i=0}^{n-1} \frac{1}{2} k_i u_i^2,
\end{align}
where $k_i$ is the staging spring constant associated with stage $i$. This form has two unknowns, for each mode, $x_i^*$ and $k_i$, which we determine in the Supporting Information. The $x_i^*$  positions are determined through first derivative of $V$ with respect to coordinates (Sec. S1 of the Supporting Information),
\begin{eqnarray}
\label{eq:xc_ho_staging}
x_i^{*}\left(x_0,x_{i-1}\right) =  
   \frac{\sinh\left(\alpha\right) { x}_0 + \sinh\left(\left(n-i\right)\alpha \right) { x}_{i-1}}{\sinh\left(\left(n-i+1\right) \alpha\right)},\, i>0
\end{eqnarray}
while the spring constants are obtained through second derivatives (Sec. S2 of the Supporting Information),
\begin{eqnarray}
\label{eq:ki_HO_staging}
k_i = \frac{m\omega_n^2}{n}\left\{
\begin{array}{ll}
      2 \sinh\left(\alpha\right) \tanh\left(\frac{n\alpha}{2}\right), & i=0 \\
       \frac{\sinh\left(\left(n-i+1\right)\alpha\right)}{\sinh\left(\left(n-i\right)\alpha\right)}, & i>0
\end{array} 
\right. 
\end{eqnarray}
where $\alpha \equiv 2\sinh^{-1}\left(\frac{\epsilon}{2}\right)$ and $\epsilon\equiv\frac{\omega}{\omega_n}=\frac{\beta\hbar\omega}{n}$. In Sec. S3 of the Supporting Information, we provide a verification on the equivalence between the original potential (Eq.~\ref{eq:V_HO_real}) and the new staging diagonalization (Eq.~\ref{eq:V_HO_staging}).

According to Eqs.~\ref{eq:ui_def} and ~\ref{eq:xc_ho_staging}, the ${\bf x} \to {\bf u}$ transformation is given by
\begin{eqnarray}
\label{eq:r2x_transformation}
u_i = x_i - \frac{\sinh\left(\alpha\right) x_0 + \sinh\left(\left(n-i\right)\alpha \right) x_{i-1} }{\sinh\left(\left(n-i+1\right) \alpha\right)}, \; i>0
\end{eqnarray}
where, again, $u_0=x_0$. The inverse (${\bf x}\leftarrow {\bf u}$) transformation is
\begin{eqnarray}
\label{eq:u2x_transformation}
x_i = u_i + \frac{\sinh\left(\alpha\right) u_0 + \sinh\left(\left(n-i\right)\alpha \right) x_{i-1} }{\sinh\left(\left(n-i+1\right) \alpha\right)}, \; i>0
\end{eqnarray}
where $x_n=x_0$, and we used that fact that $x_0=u_0$. According to Eq.~\ref{eq:r2x_transformation}, the Jacobian matrix is lower triangular, with ones along the diagonal elements. Hence, the Jacobian of the transformation (determinant) is unity. Note that, while the forward transformation is non-recursive, the inverse one is recursive (starting from $i=1$ to $n-1$). Both transformations are simple operations of element-by-element multiplications, hence the CPU cost scales with the number of beads as ${\cal O}\left(n\right)$. On the other hand, as discussed below, the NM transformations are given as matrix-vector multiplications, which yield a higher computational cost of ${\cal O}\left(n^2\right)$, or ${\cal O}\left(n\log\left(n\right)\right)$ using FFT. 

It is interesting to note that the functional form of the HO staging is identical to that of the harmonic staging associated with the PI action based on the HO reference.\cite{martyna2007low} The only difference is the $\alpha$ form, which reduces to our formula in the limit of large $n$. The similarity is attributed to the fact that both approaches diagonalize the entire HO action, with only different spring constants.

Contrary to the HO staging, the staging as known in literature (denoted here as ``FP staging") diagonalizes only the free particle (kinetic) contribution of the PI action (first term of Eq.~\ref{eq:V_PI}). Interestingly, the FP staging is a special case of the HO staging when taking the $\omega\to 0$ (or, $\alpha \to 0$) limit of Eqs.~\ref{eq:xc_ho_staging} and~\ref{eq:ki_HO_staging},
\begin{eqnarray}
\label{eq:r2x_transformation_FP}
u_i \equiv x_i-x_i^* = x_i - \frac{ x_0 +  \left(n-i\right)  x_{i-1}}{n-i+1}, & i>0
\end{eqnarray}
where $u_0=x_0$. From this, the inverse (${\bf x}\leftarrow {\bf u}$)transformation is given by
\begin{eqnarray}
\label{eq:u2x_transformation_FP}
x_i = u_i + \frac{ u_0 +  \left(n-i\right)  x_{i-1}}{n-i+1}, & i>0
\end{eqnarray}
where $x_n=x_0$. Similarly, the FP staging spring constants are given in the limit of Eq.~\ref{eq:ki_HO_staging} as
\begin{align}
\label{eq:ki_FP}
k_i &= \frac{m \omega_n^2}{n} \frac{n-i+1}{n-i},\; i>0,    
\end{align}
where $k_0=0$.

Note that we adopt a counterclockwise approach ($0,1,\dots,n-1$) to derive the HO and FP staging coordinates (see Fig. S1 in the Supporting Information). However, it is trivial to recover the clockwise version by simply replacing the $x_{i-1}$ position by $x_{i+1}$, and $n-i$ by $i$ (not including subscribes of the positions). In Sec. S4 of the Supporting Information we provide the clockwise version of both HO and FP stagings, both coordinates and spring constants. We emphasize that both versions are equivalent since they both diagonalize the same HO potential.

Figure~\ref{fig:staging_all} provides an illustrative example of a three-beads system, using both FP and HO staging methods. For the FP staging case, according to Eq.~\ref{eq:r2x_transformation_FP}, $x_i^{*}$ corresponds to an interpolation along the line connecting beads $0$ and $i-1$. On the other hand, for the HO staging method, the $x_i^{*}$ location is within the triangle connecting the origin and beads $0$ and bead $i-1$, as can be recognized from Eq.~\ref{eq:xc_ho_staging}.

\subsection{Harmonic Oscillator Normal Modes}
\label{sec:HONM}
The HO effective potential (Eq.~\ref{eq:V_HO_real}) can be expressed in terms of normal mode (NM) coordinates, $q_k$, in the following uncoupled form\cite{vath1992}
\begin{subequations}
\label{eq:eq:V_HO_nm_0}
\begin{align}
\label{eq:V_HO_nm_V}
  V\left({\bf q},\beta\right) &= \sum_{k=0}^{n-1} \frac{1}{2} \Lambda_k q_k^2, \\
  \label{eq:V_HO_nm_lambda}
\Lambda_k \equiv n\lambda_{k} &= m \omega^2 + 4m \omega_n^2 \sin^2\left(\frac{\pi k}{n}\right), k=0,1,\dots,n-1
\end{align}
\end{subequations}
where $\lambda_k$ are the eigenvalues of the following $n\times n$ Hessian matrix $H$ of second derivatives:
\begin{eqnarray}
\label{eq:hessian_ho}
H = \frac{m \omega_n^2}{n}
\begin{pmatrix}
c & -1 &  & & -1 \\ 
-1 & c & -1 & & \\ 
  &  &  \ddots & & \\ 
  & & -1&  c& -1 \\ 
  -1 & & & -1  & c 
\end{pmatrix}
\end{eqnarray}
where $c\equiv 2 + \frac{\omega^2}{\omega_n^2}$, such that $V=\frac{1}{2}{\bf x}\cdot H\cdot{\bf x}$. Note that $H$ is a symmetric circulant tridiagonal matrix.

The normal mode coordinates vector $\bf q$ is related to the Cartesian coordinates vector $\bf x$ (both of length $n$) through,\cite{markland2008NM}
\begin{align}
\label{eq:x2q_trans}
{\bf x} = \sqrt{n} A {\bf q}
\end{align}
where $A$ is a $n\times n$ matrix of the orthonormal eigenvectors associated with the Hessian matrix, hence the inverse is equal to the transpose, $A^{-1}=A^{\rm t}$. Note that the zero-frequency mode is just the centroid coordinate, $q_0=x_c$. Adopting the real-valued representation of the normal modes, the $jk$ component of $A$ is given by
\begin{eqnarray}
\label{eq:A_evecs}
A_{jk}= \frac{1}{\sqrt{n}} \left\{
\begin{array}{ll}
1, & k=0 \\
\sqrt{2}\sin\left(\frac{2\pi j k}{n}\right), & -n_2 \leq k < 0\\
\sqrt{2}\cos\left(\frac{2\pi j k}{n}\right), & 0 < k \leq n_2 \\
\left(-1\right)^j, & k=n_2+1 \;(\text{even}\; n)
\end{array} \right. 
\end{eqnarray}
where $n_2=\left(n-1\right)/2$ for odd $n$, and $n_2=\left(n-2\right)/2$ for even $n$.

A special case of the HO NM coordinates is the FP NM method, wherein only the kinetic contribution of Eq.~\ref{eq:V_HO_real} is considered in the Hessian matrix; i.e., $c=2$ in Eq.~\ref{eq:hessian_ho}. Accordingly, the FP NM eigenvalues (multiplied by $n$) are obtained by setting $\omega=0$ in Eq.~\ref{eq:V_HO_nm_lambda},
\begin{eqnarray}
\label{eq:ring_nm_lambda}
\Lambda_{k} = 4 m \omega_n^2 \sin^2\left(\frac{\pi k}{n}\right), k=0\dots n-1. \; \left(\text{FP NM}\right)
\end{eqnarray}
However, the associated eigenvectors are the \textit{same} as the HO NM case, due to the preservation of the structure of the Hessian matrix.

Note that, expressions for $x_i^{*}$ (Eq.~\ref{eq:xc_ho_staging}), $k_i$ (Eq.~\ref{eq:ki_HO_staging}), and $\Lambda_k$ (Eq.~\ref{eq:V_HO_nm_lambda}) are universal, regardless of the number of dimensions $d$ and/or atoms $N$. Hence, in the analysis below, we apply the HO staging and NM methods to sample coordinates according to the PIMC and PIMD approaches, in a general space.

\subsection{Applicability of the HO Coordinates}
\label{sec:applicability}
Below, we provide implementation details for both the PIMC and PIMD simulation methods using the HO coordinates (staging and NM). However, we first discuss the applicability domain for the proposed approach. Since the HO coordinates rely on the harmonic model (Eq.~\ref{eq:V_HO_real}), they are \textit{only} applicable to systems with a harmonic character, either explicit or implicit. Examples of the former case are systems (e.g., molecular bonds) that can be modeled explicitly with a harmonic+anharmonic potential (e.g., Eq.~\ref{eq:U_ho_ao_1d}). On the other hand, the latter case refers to systems with implicit harmonicity, i.e., through the harmonic approximation of the actual potential (e.g., using lattice dynamics\cite{dove1993book}). Examples include models for molecular bonds (e.g., Morse potential) and crystals (e.g., Lennard-Jones potential). Having said that, the method is \textit{not} applicable to fluids, systems with no harmonic contribution (e.g., purely quartic potential), or to systems with imaginary modes at the (average) equilibrium separation for molecular bonds, or at the equilibrium lattice positions for crystals.

\subsection{Application to PIMC}
\label{sec:app_pimc}
In the PIMC simulations, we sample configurations in the NVT canonical ensemble. We first introduce the sampling scheme for the HO coordinates, then present the traditional procedure for the FP coordinates case as known in literature. The HO staging and NM coordinates are sampled according to the following splitting of the total probability distribution,
\begin{align}
\label{eq:expV}
e^{-\beta V} = e^{-\beta V^{\rm HO}} \times e^{-\beta V^{\rm AH}},
\end{align}
where $V^{\rm HO}$ is the harmonic effective potential (Eq.~\ref{eq:V_HO_real}) and $V^{\rm AH}$ is the anharmonic (residual) contribution,
\begin{align}
\label{eq:Vah}
V^{\rm AH} \equiv V-V^{\rm HO} = \frac{1}{n}\sum_{i=0}^{n-1}\left(  U\left({\bf x}_i\right) - \frac{1}{2} m \omega^2 {\bf x}_i^2\right).
\end{align}
where ${\bf x}_i$ is a vector (of length $dN$) containing the beads positions associated with the $i^{\rm th}$ replica. According to this splitting, the sampling consists of two steps: a collective move proposal of the $n$ beads according to the HO contribution (Eq.~\ref{eq:V_HO_staging} for staging, or Eq.~\ref{eq:eq:V_HO_nm_0} for NM), followed by accept/reject the move according to the anharmonic contribution (Eq.~\ref{eq:Vah}). The following algorithm describes this  procedure for the HO coordinates:
\begin{enumerate}
    \item For crystals of $N$ atoms, pick a random atom, or skip to step ``2'' for quantum oscillators.
    \item For each dimension, draw $n$ transformed coordinates ($\bf u$ for staging, or $\bf q$ for NM) according to a set of Gaussian distributions with, zero means and standard deviations of
\begin{equation}
        \sigma_i = \frac{1}{\sqrt{\beta f_i}},\; i=0,\dots,n-1\nonumber
\end{equation}
where $f_i=k_i$ for HO staging (Eq.~\ref{eq:ki_HO_staging}) and $f_i=\Lambda_i$ for the HO NM (Eq.~\ref{eq:V_HO_nm_lambda}).
    \item  Transform back to the Cartesian coordinates, according to Eq.~\ref{eq:u2x_transformation} for HO staging, or Eq.~\ref{eq:x2q_trans} for HO NM methods. 
    \item  Use the standard Metropolis MC method to accept/reject the proposed move according to the \textit{change} in the anharmonic potential (Eq.~\ref{eq:Vah}) between the old and new configurations, $\Delta V^{\rm AH}$.
    \item If the move is accepted, update the beads positions, otherwise use the old configuration.
    \item Repeat.
\end{enumerate}
For the case of crystals, the displacements from the lattice sites are used as the Cartesian coordinates.

This algorithm is applicable to the FP coordinates as well, with the following exceptions. First, in step 2, the non-zero collective modes are still sampled according to Eq.~\ref{eq:V_HO_staging} for staging and Eq.~\ref{eq:eq:V_HO_nm_0} for NM, but using the FP force constants; i.e., Eq.~\ref{eq:ki_FP} for staging and Eq.~\ref{eq:ring_nm_lambda} for NM. This move results in a change in the center of mass (centroid), which we bring back to where it was after the move. Second, the centroid (zero-mode) is then sampled separately using a rigid translation move of the ring-polymer (with a target acceptance of $50\%$). Both of these moves are proposed with an equal probability. Third, the inverse transformation (step 3) for the FP staging is done using Eq.~\ref{eq:u2x_transformation_FP}. But same transformation is used for both HO and FP coordinates (Eq.~\ref{eq:x2q_trans}). Lastly, the residual energy used in step 4 is the difference from the kinetic term
\begin{equation}
\label{eq:V_residual_RP}
V^{\rm residual}\equiv V-V^{\rm FP}=\frac{1}{n}\sum_{i=0}^{n-1} U\left({\bf x}_i\right),    
\end{equation}
which is just the external energy contribution.

\subsection{Application to PIMD}
Sampling configurations in PIMD is governed by fictitious dynamics, according to the following effective Hamiltonian in a general (staging or NM) phase space of coordinates $q$ and momenta $p$:
\begin{align}
H\left(q,p\right) = \sum_l \frac{p_l^2}{2m_l} + V\left(q,\beta\right),
\end{align}
where $m_l$ is a fictitious mass (defined below) associated with the $l$ degree of freedom and $V$ is the effective PI potential (Eq.~\ref{eq:V_HO_staging} for staging and Eq.~\ref{eq:V_HO_nm_V} for NM). We run PIMD simulations in NVT canonical ensemble using Langevin thermostat, which we detail next.  

\subsubsection{White noise Langevin thermostat}
The Langevin dynamics equations of motion, in a general coordinate system, can be split into three parts,\cite{Leimkuhler2013,allen2017book}
\begin{eqnarray}
\label{eq:pimd_splitting}
\begin{pmatrix}
 \dot{q}_l \\
 \dot{p}_l
\end{pmatrix}
=
\underbrace{\begin{pmatrix}
 \frac{p_l}{m_l} \\
0
\end{pmatrix}}_\text{A}
+
\underbrace{\begin{pmatrix}
 0 \\
F_l
\end{pmatrix}}_\text{B}
+ 
\underbrace{\begin{pmatrix}
 0 \\
-\gamma p_l + \sqrt{2m_lk_{\rm B}T\gamma} \; \xi_l
\end{pmatrix}}_\text{O}
\end{eqnarray}
where, for a given $l$ degree of freedom, $F_l\equiv-\partial V/\partial q_l$ is the effective force, $\xi_l$ describes the Gaussian white noise, with $\left<\xi_l\left(t\right)\right>=0$ and $\left<\xi_l\left(t\right)\xi_{l'}\left(t'\right)\right>=\delta_{ll'} \delta\left(t'-t\right)$, where $\delta_{ll'}$ the Kronecker delta function, and $\gamma$ is the thermostat friction coefficient (set to $\gamma=\omega$ for optimum performance\cite{liu2016simple}). The ``A'' part is a free flight drift, while ``O'' (Ornstein-Uhlenbeck process) represents random kicks in momenta required to guarantee a canonical ensemble, both are solved exactly. On the other hand, the ``B'' contribution represents deterministic kicks in momenta, caused by the total force $F_l$, and it sets an upper bound on the Verlet time step size.\cite{Leimkuhler2013} 

We point out that a more common splitting approach (FP coordinates) involves using the ring-polymer forces with ``A'' and the physical forces with ``B''.\cite{ceriotti2010efficient,liu2016simple} In this case, the ``A'' propagator represents a harmonic system, which is solved exactly. We explored this option in Sec. S6 of the Supporting Information. However results show that the performance of our splitting choice is still better, in terms of convergence of properties with time step size. The same observation is already reported by Liu et al., where they denoted the splitting we adopted as ``BAOAB-num''.\cite{liu2016simple} Moreover, other studies show that propagating the ``A'' part approximately, using Cayley transformation, can avoid numerical instabilities.\cite{korol2019cayley,korol2020dimension} We also considered this option in the Supporting Information and results still supports our splitting choice in Eq.~\ref{eq:pimd_splitting}
 
Due to its efficiency, we adopt the BAOAB integrator scheme, in which a half time step is used to propagate the positions (A) and momenta (B), and a full time step for the random kick (O),\cite{Leimkuhler2013,liu2016simple}
\begin{subequations}
\label{eq:baoab}
\begin{align}
{\rm B}\left(\frac{\Delta t}{2}\right): \;  &p_l \leftarrow p_l  + F_l \frac{\Delta t}{2}  \\
{\rm A} \left(\frac{\Delta t}{2}\right): \; &q_l \leftarrow q_l  + \frac{p_l}{m_l} \frac{\Delta t}{2}\\
{\rm O} \left(\Delta t\right): \; &p_l \leftarrow p_l \exp\left(-\gamma \Delta t\right) +  \sqrt{m_l k_{\rm B}T\left(1-\exp\left(2\gamma\Delta t\right)\right)} \; R_l\\
{\rm A} \left(\frac{\Delta t}{2}\right): \; &q_l \leftarrow q_l  + \frac{p_l}{m_l} \frac{\Delta t}{2} \\
{\rm B} \left(\frac{\Delta t}{2}\right): \; &p_l \leftarrow  p_l + F_l \frac{\Delta t}{2}
\end{align}
\end{subequations}
where $R_l$ is a normal distribution random number, with $\left<R_l\right>=0$ and $\left<R_l R_{l'}\right>=\delta_{ll'}$. For completeness, we also considered the OBABO ordering option,\cite{ceriotti2010efficient,korol2019cayley,liu2016simple} and reported results in Sec. S7 of the Supporting Information. In all cases considered, the OBABO scheme shows a poor dependence on the time step size, relative to the current BAOAB approach, which is an observation that has already been reported by others.\cite{liu2016simple,korol2020dimension}

We apply this algorithm to PIMD, in both staging and NM coordinates (HO and FP methods). In this application, the system evolves in the transformed coordinates according to the BAOAB procedure above. However, since the actual forces are defined in the Cartesian coordinates, we first perform inverse transformation of the coordinates (${\bf q}\rightarrow {\bf x}$), followed by force evaluation in the Cartesian space. These forces are then transformed to the staging/NM coordinate system, which is discussed in the next two sections for the staging and NM coordinates.

\subsubsection{Cartesian to staging force transformation}
For a given atom, the force acting on bead $j$ in the HO or FP staging coordinates is given by, 
\begin{eqnarray}
\label{eq:eom_stage}
F_j =  -\frac{\partial V}{\partial u_j}, \; j=0,1,\dots n-1
\end{eqnarray}
Using the chain rule, these forces can be represented in terms of the Cartesian forces, $F_j^{xyz}= - \partial V/\partial x_i $, as
\begin{eqnarray}
\label{eq:F_s2r_0}
F_j = \sum_{i=j}^{n-1}  \frac{\partial x_i}{\partial  u_j} F_i^{xyz},
\end{eqnarray}
where $F^{xyz}_i= - \frac{m \omega_n^2}{n} \left(2 x_i - x_{i-1} - x_{i+1} \right)   -\frac{1}{n} \frac{\partial U\left(x_i\right)}{\partial x_i}$, according to Eq.~\ref{eq:V_PI}. To evaluate Eq.~\ref{eq:F_s2r_0}, we need a closed form of the coordinate transformation, which can be obtained recursively using Eq.~\ref{eq:u2x_transformation},
\begin{align}
x_i\left(u_0,\dots,u_i\right) &=
\frac{\cosh\left(\left(\frac{n}{2}-i\right)\alpha\right)}{\cosh\left(\frac{n}{2}\alpha\right)} u_0 + \sum_{j=1}^{i} \frac{\sinh\left(\left(n-i\right)\alpha\right)}{\sinh\left(\left(n-j\right)\alpha\right)} u_j
\end{align}
Using this in Eq.~\ref{eq:F_s2r_0}, yields the following backward recursion relations for forces,
\begin{align}
\label{eq:F_r2s}
    F_j = \left\{
\begin{array}{ll}
   \sum_{i=0}^{n-1}  \frac{\cosh\left(\left(\frac{n}{2}-i\right)\alpha\right)}{\cosh\left(\frac{n}{2}\alpha\right)} F_i^{xyz},&j=0 \\
       F_j^{xyz}
+ \frac{\sinh\left(\left(n-j-1\right)\alpha\right)}{\sinh\left(\left(n-j\right)\alpha\right)}
F_{j+1},&j=n-1,\dots, 1
\end{array} 
\right. & \nonumber \\ 
(\text{HO staging})  &
\end{align}
with $F_n=F_0$.

On the other hand, the FP staging forces are given as a special case of the HO staging  $\omega\to 0$ (or, $\alpha\rightarrow 0$)  limit in Eq.~\ref{eq:F_r2s}, 
\begin{align}
    F_j &= \left\{
\begin{array}{ll}
   \sum_{i=0}^{n-1}  F^{xyz}_i, & j=0 \\
       F^{xyz}_j
+ \frac{ n-j-1 }{n-j }
F_{j+1}, &  j=n-1,\dots, 1
\end{array} 
\right.     (\text{FP staging})  
\end{align}
Note that, this transformation is already known in literature,\cite{tuckerman1993efficient,liu2016simple} but using the clockwise staging (see Sec. S4 of the Supporting Information).

\subsubsection{Cartesian to NM force transformation}
Given that the eigenvectors for the NM coordinates using both the HO and FP methods are the same, it follows that the corresponding force transformations are also identical. Due to the nature of the NM transformations (Sec.~\ref{sec:HONM}), we adopt a matrix notation throughout. Similar to the staging case, we will focus on the beads coordinates associated with a given atom. The forces acting on the $n$ beads, in the NM coordinates $\bf q$, are given by the following force vector,
\begin{align}
{\bf F} = -\frac{\partial V}{\partial {\bf q}} 
\end{align}
where both $\bf q$ and $\bf F$ vectors are of length $dn$. Using the chain rule, along with the ${\bf q}\to {\bf x}$ transformation (Eq.~\ref{eq:x2q_trans}), theses forces are expressed in terms of the Cartesian coordinates forces ${\bf F}^{xyz}$ as follows:
\begin{eqnarray}
\label{eq:force_trans_nm}
{\bf F} =  \frac{\partial {\bf x}}{\partial {\bf q}} \cdot {\bf F}^{xyz}  
= \sqrt{n} A^t {\bf F}^{xyz} 
\end{eqnarray}
where ${\bf F}^{xyz}\equiv-\frac{\partial V}{\partial {\bf x}}$.  

It is worthwhile to emphasize that, despite the simplicity of the NM transformations, the computational cost of these force transformations is of order ${\cal O}\left(n^2\right)$, or ${\cal O}\left(n\log\left(n\right)\right)$ using FFT, due to the matrix-vector multiplication operations. This differentiates it from the staging method, which is only of order ${\cal O}\left(n\right)$, because of the recursive nature of the transformations (e.g., Eq.~\ref{eq:F_r2s}).

\subsubsection{Fictitious masses}
\label{sec:fic_masses}
Efficient PIMD simulations narrow the gap between the diverse time scales of PI systems by introducing a set of mode-based fictitious (or dynamic) masses.\cite{muser2002} This freedom of choosing the masses allows for using larger MD time steps, compared to the small time steps needed for the stiff internal PI vibrations. We define these masses in the HO staging and NM coordinates such that all modes (including the centroid) fluctuate on the \textit{same} time scale for HO systems. However, for a meaningful comparison across different coordinates at a given time step $\Delta t$, we require this time scale to match the motion of the centroid in the Cartesian space. We set the physical mass of each bead to $m/n$; hence, the centroid has a mass of $m$ and oscillates at a frequency of $\omega$. To achieve this with the HO methods, we define the fictitious masses as $m_i\equiv k_i/\omega^2$ for the HO staging (Eq.~\ref{eq:ki_HO_staging}),
\begin{eqnarray}
m_i = \frac{m\omega_n^2}{n\omega^2} \left\{
\begin{array}{ll}
      2  \sinh\left(\alpha\right) \tanh\left(\frac{n\alpha}{2}\right),i=0 \\
         \frac{\sinh\left(\left(n-i+1\right)\alpha\right)}{\sinh\left(\left(n-i\right)\alpha\right)},i=1,2,\dots,n-1
\end{array} 
\right.(\text{HO staging}) 
\end{eqnarray}
and as $m_k\equiv \Lambda_k/\omega^2$ for the HO NM (Eq.~\ref{eq:V_HO_nm_lambda}), 
\begin{eqnarray}
m_k = m \left[ 1 + 4  \frac{\omega_n^2}{\omega^2} \sin^2\left(\frac{\pi k}{n}\right) \right], k=0,1,\dots,n-1  (\text{HO NM})  
\end{eqnarray}
Since these choices unify the time scales for the HO model, anharmonicity would disturb this unity and could affect the allowed time step size, hence the sampling performance. However, still the performance is expected to be efficient relative to the FP methods, as supported by current results. These expectations are further discussed in the Results sections in for the AO model.

On the other hand, for the FP coordinates, we adopt the fictitious masses definitions as already given in the literature.\cite{tuckerman1996efficient,witt2009,liu2016simple} For FP staging and NM coordinates, these masses are given, respectively, by
\begin{align}
\label{eq:mi_ring_stage}
    m_i &= m\left\{
\begin{array}{ll}
   1, & i=0 \\
       \frac{n-i+1}{n-i}, & i>0
\end{array} 
\right.     (\text{FP staging})  
\end{align}
and
\begin{align}
\label{eq:mi_ring_nm}
m_k &= m\left\{
\begin{array}{ll}
     1, & k=0 \\
     4 n \sin^2\left(\frac{\pi k}{n}\right) , & k>0
\end{array} 
\right.      (\text{FP NM})    
\end{align}
Note that, for the FP NM case, the zero-mode represents the centroid coordinate; i.e., $q_0=x_{\rm c}$, see Sec.~\ref{sec:HONM}. Therefore, the $m_0=m$ definition already yields a centroid oscillation of a frequency $\omega$ in the real space, which satisfies the aforementioned criterion of matching the centroid motion. Unlike the NM case, all the staging $u_i$ modes contribute to the centroid, which results in a complex dynamics of its motion. However, since the zero staging mode, $u_0$, defines the external Cartesian (absolute) coordinate of one bead, it is expected to somewhat resemble the centroid motion. In fact, we find this to be the case for temperatures above some intermediate value ($\approx 0.4$), using PIMD simulations in NVE ensemble (not shown). This is to be expected as the internal interactions become stiffer with increasing the temperature (large $n$), such that their contribution to the centroid becomes smaller in comparison to $u_0$. Thus, the $m_0=m$ choice with FP staging also yields (yet approximately) a centroid oscillation with a frequency $\omega$. However, at low temperatures, the centroid oscillates faster with complex dynamics, such that comparing performance with other methods, at a given $\Delta t$, not meaningful. However, in practice, interest is in determining the maximum allowable time step size that preserves accuracy.

\subsection{Centroid Virial Estimator}
We consider the total energy and heat capacity at constant volume as example properties to assess the sampling efficiency of both PIMC and PIMD simulations. We adopt the centroid virial estimator for both quantities, due to its known computational advantages over others  (e.g., thermodynamic or virial estimators).\cite{shiga2005cv} For a $d$-dimensional system of $N$ species and $n$ beads, the average total energy $E$ and heat capacity $C_{\rm V}$ are given, respectively, as
\begin{subequations}
\begin{align}
 \label{eq:E_cvir}
E  &= \frac{dN}{2\beta} + \left<  \sum_{i=0}^{n-1} \frac{1}{n}  U\left({\bf x}_i\right)   - \frac{1}{2}  \sum_{i=0}^{n-1}  {\bf F}_i^{\rm phy} \cdot   \left( {\bf x}_i - {\bf x}_c\right)  \right>, \\
\label{eq:Cv_cvir}
\frac{C_{\rm V}}{k_{\rm B} \beta^2}   
&=  \frac{d N}{2\beta^2}  +    {\rm Var}\left(E\right) + 
\frac{1}{4\beta} \left< 3  \sum_{i=0}^{n-1}  {\bf F}_i^{\rm phy}  \cdot  \left( {\bf x}_i - {\bf x}_c\right) \right. \nonumber\\
&- \left.   \sum_{i=0}^{n-1} \left( {\bf x}_i - {\bf x}_c\right) \cdot  H_{ii}^{\rm phy} \cdot \left( {\bf x}_i - {\bf x}_c\right) \right>,
 \end{align}   
\end{subequations}
where ${\bf x}_c \equiv \frac{1}{n}\sum_{i=0}^{n-1} {\bf x}_i$ is a vector of the centroid coordinates associated with each atom (of length $dN$). Moreover, the physical forces (of length $dN$) and Hessian matrix ($dN\times dN$) associated with the $i^{\rm th}$ replica are given in terms of the intermolecular potential $U$ as ${\bf F}^{\rm phy}_i=-\frac{1}{n}\frac{\partial U\left({\bf x}_i\right)}{\partial {\bf x}_i}$ and $H_{ii}^{\rm phy}=\frac{1}{n}  \frac{\partial^2 U\left({\bf x}_i\right)}{\partial {\bf x}_i \partial {\bf x}_i}$, respectively. The superscript ``0'' is used to denote the absence of internal harmonic interactions (i.e., kinetic term in Eq.~\ref{eq:V_PI}). Finally, ${\rm Var}\left(E\right)\equiv \left<E^2\right>-\left<E\right>^2$ is the variance of the instantaneous energy (quantity being averaged in Eq.~\ref{eq:E_cvir}).

\subsection{Uncertainty Analysis}
Uncertainties (correspond to a 68\% confidence limits) in the energy (Eq.~\ref{eq:E_cvir}) and heat capacity (Eq.~\ref{eq:Cv_cvir}) are estimated from single runs, using the block averaging technique.\cite{book2023understanding} This is straightforward to implement with the energy, since it is given as an average. However, the heat capacity is given as a variance and, hence, propagation of uncertainty is needed to express its uncertainty in terms of measured averages. According to Eq.~\ref{eq:Cv_cvir}, the heat capacity can be written in this compact form:    
\begin{eqnarray}
C_{\rm V} = \bar Y -  {\bar E {\Large\mathstrut}}^2,
\end{eqnarray}
where ${\bar E {\large\mathstrut}}^2$ is the square of the block averaged energy, and $\bar Y$ is a block average of the remaining terms. Propagating the uncertainty to the second-order in fluctuations, yields 
\begin{eqnarray}
\sigma^2\left(C_{\rm V}\right) 
= \sigma^2\left(\bar Y\right) 
      + 4 {\bar E {\Large\mathstrut}}^2 \sigma^2\left(\bar E\right)
     - 4 {\bar E} \, {\rm cov}\left({\bar Y},{\bar E}\right) 
\end{eqnarray}
where $\sigma^2\left(Y\right)$ and ${\rm cov}\left(Y,Z\right)$ are the variance and covariance, respectively. For a sufficiently large block size, the blocks can be considered independent, and the uncertainty can be written in terms of the statistics of the raw data, 
\begin{eqnarray}
\sigma^2\left(C_{\rm V}\right)  =  \frac{1}{N_{\rm blocks}}\left[ \sigma^2\left(Y\right) 
      + 4 {\bar E {\Large\mathstrut}}^2 \sigma^2\left(E\right) 
     - 4 {\bar E} \, {\rm cov}\left(Y,E\right)\right]    
\end{eqnarray}
where $N_{\rm blocks}$ is the number of blocks ($100$ in our case). We verified the correctness of this formula against direct measurement of the standard deviation from $100$ independent runs.

\subsection{Simulation Details}
Two, one-dimensional, quantum oscillators are considered as prototypes: a harmonic oscillator (HO) and an asymmetric anharmonic oscillator (AO), with the following potentials,
\begin{subequations}
\label{eq:U_ho_ao_1d}
\begin{align}
\label{eq:U_ho_1d}
U^{\rm HO} &= \frac{1}{2} m\omega^2 x^2, \\
\label{eq:U_ao_1d}
U^{\rm AO} &=\frac{1}{2}m\omega^2 x^2  +k_3 x^3 + k_4 x^4,        
\end{align}
\end{subequations}
where we set $m=\omega=1$, $k_3=0.1$, and $k_4=0.1$ throughout this work. Despite their simplicity, these models have all the problems associated with PI simulations (e.g., nonergodicity, multiple time scales, CPU scaling with $n$, etc.). In addition, the HO has an exact energy form in the primitive PI action of $n$ beads,\cite{schweizer1981,cao1993born,chin2023analytical} 
\begin{align}
\label{eq:En_ho_exact}
E = \frac{\hbar\omega }{2}\frac{\coth\left(\frac{n\alpha}{2}\right)}{\sqrt{1+\frac{1}{4}\epsilon^2}} 
\end{align}
where $\alpha$ and $\epsilon$ are as defined above. The corresponding energy in the continuum limit ($n\to \infty$, or $\epsilon\to 0$) is then given by,
\begin{equation}
\label{eq:E_ho_inf}
E_{\infty} = \frac{\hbar\omega}{2} \coth\left(\frac{\beta\hbar\omega}{2}\right). \text{(continuum limit)} 
\end{equation}
The heat capacity at constant volume, $C_{\rm V}$, can then be obtained from the first derivative of the energy with respect to temperature. For the AO model, on the other hand, we use the numerical matrix multiplication (NMM) method\cite{NMM1983,NMM2001} to get exact Helmholtz free energy, from which we determine the total energy and heat capacity using a five-point finite difference scheme.

\begin{figure}
\includegraphics[width=0.35\textwidth]{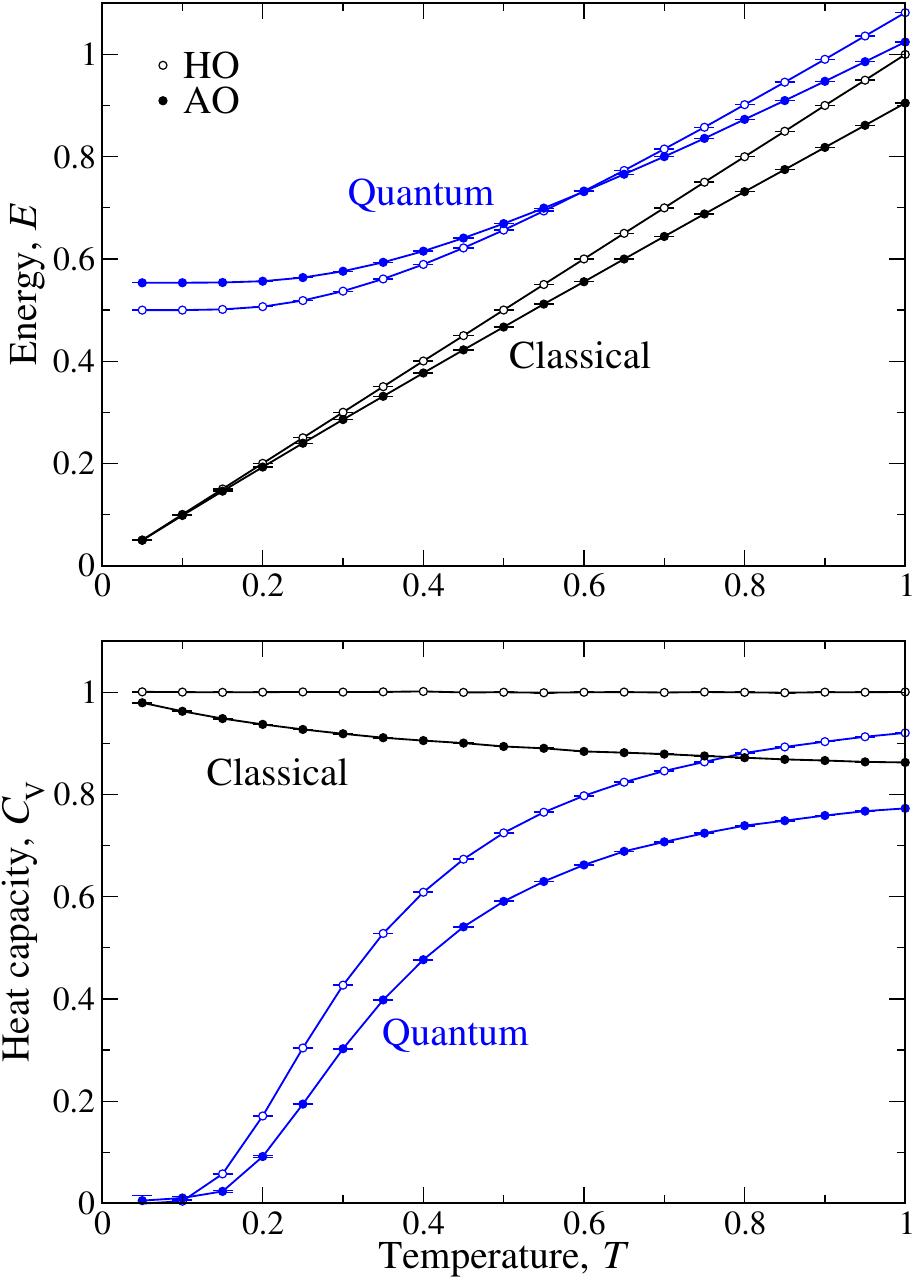}
\caption{Temperature dependence of the energy $E$ (top) and heat capacity at constant volume $C_{\rm V}$ (bottom) of a one-dimensional HO and AO models (Eqs.~\ref{eq:U_ho_ao_1d}). The classical treatment ($n=1$) is presented to show the level of quantumness. Error bars correspond to a 68\% confidence limits, here and throughout this work. Most error bars are smaller than the symbol size and lines join the data points as a guide to the eye.  The data were generated from PIMC simulations, using the HO staging coordinates.}
\label{fig:E_Cv_T}
\centering
\end{figure}

Both PIMC and PIMD simulations ran in NVT canonical ensemble, for $10^7$ steps, after $10^6$ steps of equilibration. A temperature range of $T=0.05-1.0$, with an increment of $0.05$, was considered, which shows significant quantum effects (see Fig.~\ref{fig:E_Cv_T}). For the PIMD simulations, the time step size was chosen for each method to be the largest possible value that did not sacrifice accuracy. Specific values for each case are given in the Results section. White noise Langevin (stochastic) thermostat was adopted with PIMD to ensure a canonical distribution, using the BAOAB integrator method.\cite{Leimkuhler2013,liu2016simple} The thermostat friction coefficient was set to $\gamma=\omega$, which is the common frequency of all modes in the HO staging and NM coordinates. Although this is the optimum value in the absence of external field,\cite{liu2016simple} current (and previous\cite{liu2014path}) results did not show sensitivity to $\gamma$ in the vicinity of this value.

Based on the convergence analysis with respect to the number of beads, we find that $n=20 \beta\hbar\omega$ to yield statistically converged energy and heat capacity, for both the HO and AO models (see Sec. S5 of the Supporting Information). However, in order to compare against the exact NMM method (where $n$ is multiples of $2$), we require $n=2^i$, where $i$ is the smallest integer number such that $n\ge 20 \beta\hbar\omega$. This condition, for example, corresponds to $n=32$ ($T=1.0$) and $256$ ($T=0.1$). It is worthwhile to emphasize that the scaling factor ($20$) is only suitable for the level of precision attained here --- larger $n$ values would be needed for higher precision levels.

Both PIMC and PIMD simulations were performed using the Etomica simulation software,\cite{schultz2015etomica} which can be accessed at https://github.com/etomica/etomica/tree/path\_integral. 

\section{RESULTS AND DISCUSSION}
\label{sec:results}
To gain an insight on the degree of quantumness within the given temperature range ($T=0.05-1.0$), we present in Fig.~\ref{fig:E_Cv_T} the quantum and classical ($n=1$) representations of the total energy and heat capacity, for both the HO and AO models. The data were generated using PIMC simulation, in HO staging coordinates. We notice that the difference between the quantum and classical representations is substantial across the whole temperature range considered, with the maximum value at zero temperature (i.e., zero-point energy). We now use these highly quantum models to investigate the performance of the staging and NM coordinates, using both PIMC and PIMD simulation methods.

\subsection{PIMC}   
\begin{figure}
\includegraphics[width=0.49\textwidth]{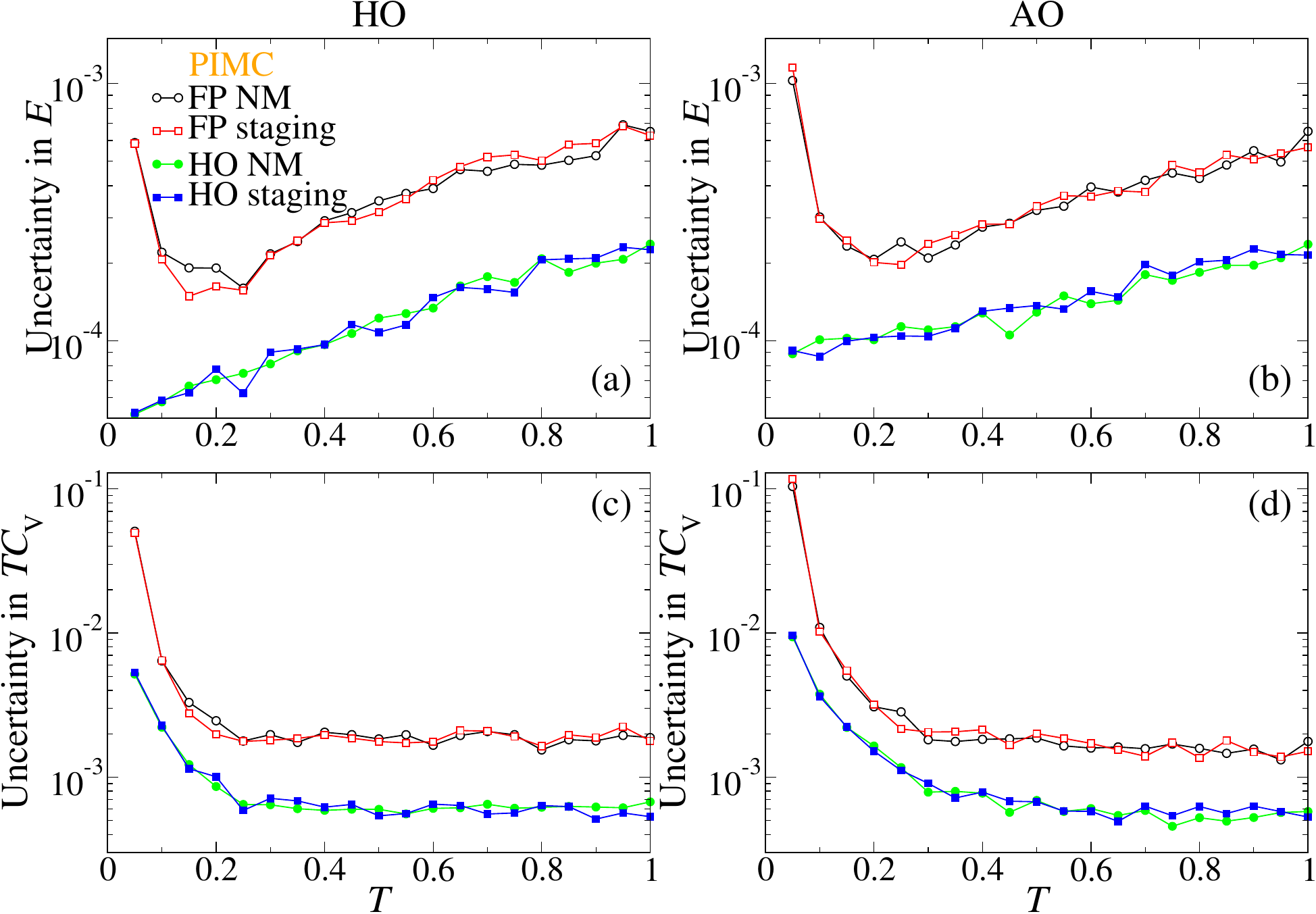}
\caption{Temperature dependence of the PIMC statistical uncertainty in estimating the energy (top) and heat capacity multiplied by the temperature (bottom) of the HO (left) and AO (right) models, using both the HO and FP coordinates.}
\label{fig:err_T_pimc}
\centering
\end{figure}

Figure~\ref{fig:err_T_pimc} presents the PIMC statistical uncertainty in estimating the energy (top) and heat capacity multiplied by $T$ (bottom) of the HO model, using both HO and FP methods. Results from the AO model (not shown) are qualitatively similar. The first observation to report is the statistically indistinguishable performance of the staging and NM coordinates, for a given FP/HO diagonalization method. This is to be expected because both coordinates diagonalize the same energy function; i.e., the kinetic contribution of Eq.~\ref{eq:V_HO_real} for the FP method and the total for HO. The figure also shows that uncertainties in both $E$ and $C_{\rm V}$ using the HO method are consistently smaller than those of the FP methods --- about $10\times$ and $3\times$ smaller at the low and high temperatures, respectively. Thus, for a given number of MC steps, the HO method generates more independent samples than the FP approach. This could be attributed the nature of both MC moves. While the HO method samples all degrees of freedom at once, including the centroid, the FP approach uses a separate translation move for the centroid. Therefore, the FP coordinates need more steps to reach to an independent sample. In addition, the trial potential associated with the collective move of the HO staging (Eq.~\ref{eq:Vah}) is different from that of FP staging (Eq.~\ref{eq:V_residual_RP}), which improves the acceptance rates for HO. The combined effect of these factors yield the lower uncertainties observed with the HO coordinates. 

A closely related observation is that uncertainties in $E$ and $C_{\rm V}$ using the FP methods increase substantially at low temperatures (below $T\approx 0.2$), while the behavior of the HO methods is smooth and consistent in the whole temperature range. This could be understood in the context of the temperature dependence of the size of the ring-polymer. While the actual size at low temperatures (including $T=0$) is finite, the corresponding size of the free particle model (used in the FP methods) is given by the thermal de Broglie wavelength,\cite{ceperley1995review} which diverges as the temperature approaches zero. Therefore, the proposed FP moves are not suitable at low temperatures, resulting in low acceptance rates and, hence, higher uncertainties. Different ``fixes'' are often adopted to enhance the sampling in the FP coordinates at low temperatures (e.g., partial FP sampling\cite{tuckerman1993efficient}). On the other hand, no such fixes are needed with the HO coordinates, which make them simpler to implement and, yet, more efficient.

 \subsection{PIMD}
\begin{figure}
\includegraphics[width=0.48\textwidth]{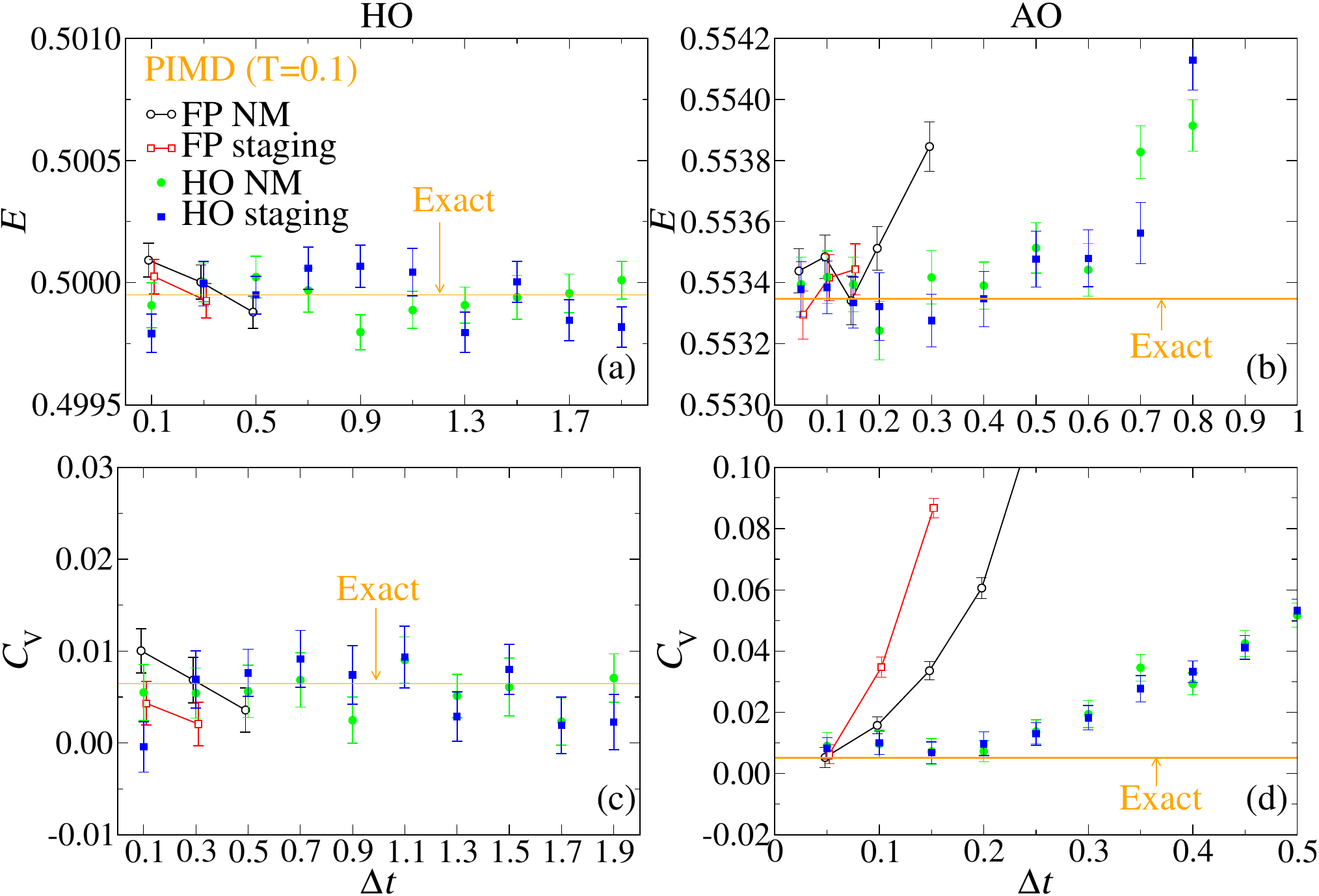}
\centering
\caption{Dependence of the energy (top) and heat capacity (bottom) on the PIMD time step size, for both harmonic (left) and anharmonic (right) oscillators, at $T=0.1$. The horizontal orange lines represent exact values, using Eq.~\ref{eq:En_ho_exact} for the HO model and the NMM method for the AO model, at the given number of beads ($n=256$). The number of steps $N_{\rm steps}$ is chosen such that the simulation time is kept fixed for all points, $t_{\rm sim}=N_{\rm steps} \Delta t=10^7$. The FP data are shifted slightly left/right for clarity. Presented results for the FP coordinates correspond to stable integrator (simulations failed for larger $\Delta t$ values).}
\label{fig:dt_conv}
\end{figure}

\begin{figure}
\includegraphics[width=0.48\textwidth]{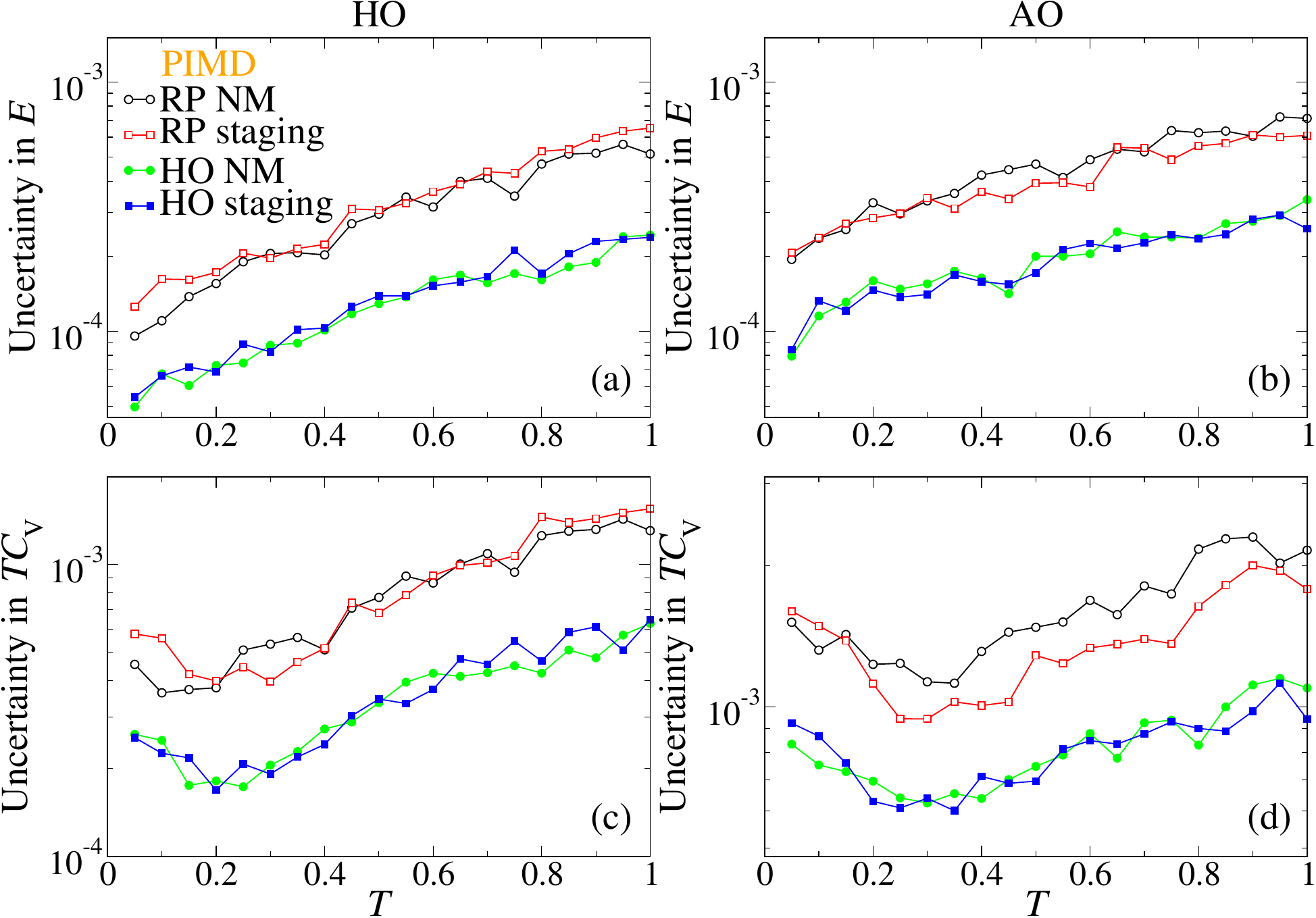}
\caption{Same as Fig.~\ref{fig:err_T_pimc}, but using PIMD. For each data set, we choose the largest time step possible that yields accurate results (across all temperatures) compared to the exact value, while fixing the number of steps to $N_{\rm steps}=10^7$. For the HO model, the $\Delta t$ values (both $E$ and $C_{\rm V}$) are $0.3$ for FP NM, $0.2$ for FP staging, and $1.8$ for HO coordinates. For the AO model, the values for $E$ are $0.1$ and $0.6$ for the FP and HO coordinates, respectively, while the values for $C_{\rm V}$ are $0.05$ and $0.2$ for the FP and HO coordinates, respectively. }
\label{fig:err_T_pimd}
\centering
\end{figure}

Here, we investigate the PIMD sampling efficiency based on the effect of time step size (accuracy) and statistical uncertainty (precision). Figure~\ref{fig:dt_conv}, shows the effect of time step on the energy (top) and heat capacity (bottom), for both HO (left) and AO (right) models. Results are shown for $T=0.1$; however, a similar behavior is observed at other temperatures. To ensure nearly similar statistical uncertainties from all time steps, we fix the simulation time to $t_{\rm sim}=10^7$. We first recall that the  fictitious masses used correspond to same oscillation ($\omega$) of the centroid motion, such that comparing different methods at the same $\Delta t$ is meaningful. For the HO model, the HO coordinates provide statistically exact solution, for both properties, with a maximum time step size corresponds to the velocity Verlet threshold value ($\Delta t=2/\omega=2.0$). This observation is already established analytically for the HO model, using the same BAOAB velocity Verlet integrator.\cite{Leimkuhler2013} On the other hand, substantially smaller values ($\Delta t < 0.7$) are required for the FP method. This is a manifestation of the fact that the FP coordinates only diagonalize the FP contribution and, hence, they do not efficiently narrow the gap between multiple time scales. Although anharmonicity (right column) negatively impacts performance of all methods, results from HO coordinates are still the most efficient. For the energy, while the HO coordinates allow time steps as large as $\Delta t\approx 0.6$, the FP coordinates only tolerate sizes up to $\approx 0.15$. The heat capacity, on the other hand, shows more sensitivity to $\Delta t$, yet the HO coordinates appear to handle larger values ($\approx 0.2$) than the FP coordinates ($\approx 0.05$). As mentioned earlier, to provide a fair comparison to the FP coordinates method, we also consider other integrator splitting and ordering (OBABO) in the Supporting Information, However, results show that the performance or our integrator scheme (Eqs.~\ref{eq:pimd_splitting} and~\ref{eq:baoab}) performs better, or at least similar to those alternative in some cases.

Moreover, for a given model and property, the size of the error bars (68\% confidence limit) from all methods appear to be statistically similar. To explain this, we first recognize the fact that the statistical uncertainty depends on the standard deviation (independent of $\Delta t$) and the number of independent samples. For a given simulation time, the number of independent samples depends on the slowest time scale present in the system, which corresponds here to the centroid mode. Therefore, since we force the centroid motion to match across all methods, the number of independent samples is expected to be similar, which yields the similar uncertainties we observe.

The direct effect of having large time step size, for a given $t_{\rm sim}$, is to improve the precision in measured properties. Figure~\ref{fig:err_T_pimd} depicts the temperature-dependence of the statistical uncertainty in estimating the energy (top) and heat capacity multiplied by temperature (bottom) of the asymmetric AO model. Note that the PIMD simulations ran using the largest time step sizes possible for \textit{all} temperatures (given in the figure caption). The accuracy in measured $E$ and $C_{\rm V}$ using these choices was confirmed using the exact analytical solution of the HO model and using the exact NMM method\cite{NMM1983,NMM2001} for the AO model. The relative performance of the HO and FP methods is similar to that of the PIMC case, with the exception that the FP data do not diverge at low temperatures. This could be explained by the same aforementioned argument that the number of independent samples depends on the longest time scale, which is forced to match across all methods, regardless of the temperature. Hence, the number of independent samples is nearly constant, such that the variation of precision with temperature depends primarily on the standard deviation (a smooth function of $T$).

\subsection{Asymmetric Anharmonic Effects}

\begin{figure}
\includegraphics[width=0.48\textwidth]{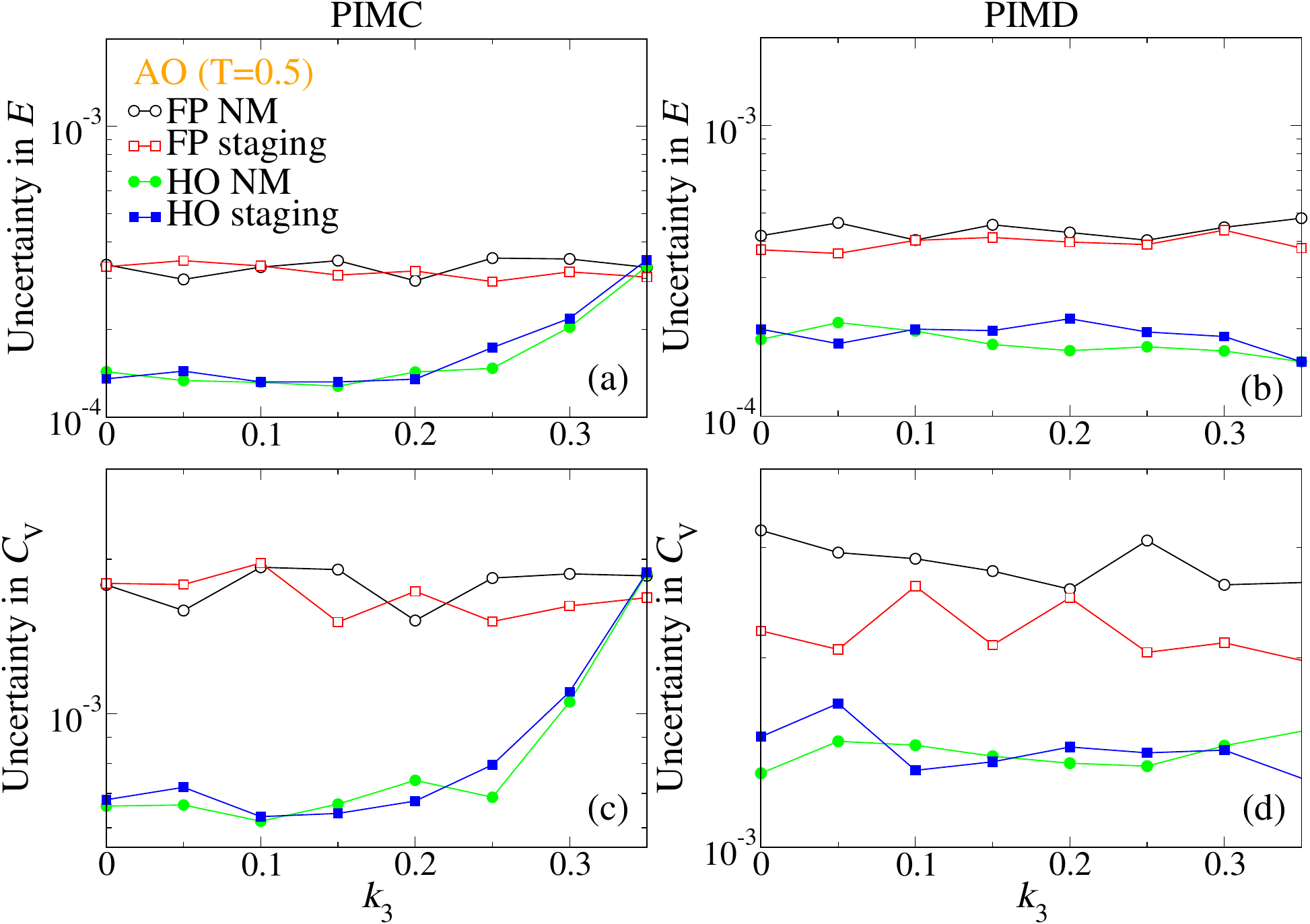}
\centering
\caption{Dependence of the statistical uncertainty in the energy (top) and heat capacity (bottom), using PIMC (left) and PIMD (right) at $T=0.5$, on the asymmetric spring constant, $k_3$. The quartic spring constant is kept fixed to $k_4=0.1$. The same $\Delta t$ values as in Fig.~\ref{fig:err_T_pimd} are used to generate the PIMD data, which provide accurate estimates.}
\label{fig:err_k}
\end{figure}

Here, we study the effect of asymmetry of the AO model ($k_3$, Eq.~\ref{eq:U_ao_1d}) on the sampling efficiency. In terms of the potential energy surface, increasing $k_3$ (for a given $k_2$ and $k_4$) widens the left side of the potential, while steepening the right side. However, a second minimum will develop at a value of $k_3^*=\frac{4}{3}\sqrt{k_4}$, which corresponds to $k_3\approx 0.4216$ in our case. Clearly, our HO coordinates method will fail beyond this point; hence, we restrict ourselves to smaller values. Figure~\ref{fig:err_k} depicts the effect of this asymmetry on the precision of estimating the energy (top) and heat capacity (bottom), using the PIMC (left) and PIMD (right) methods. The data are shown for an intermediate temperature ($T=0.5$), however a qualitatively similar behaviour is observed at other temperatures. For PIMD, the relative performance of the HO and FP coordinates methods do not appear to be sensitive to anharmonicity, for both $E$ and $C_{\rm V}$. For PIMC, on the other hand, the HO coordinates provide more precise estimates at small to moderate $k_3$ values; however, the improvement reduces as the asymmetry increases towards $k_3^*$. The difference in the effect of asymmetry on the performance of the PIMC and PIMD methods could be attributed to the difference in the effect of coordinates on each method.

\subsection{Computational Cost}
\begin{figure}
\includegraphics[width=0.35\textwidth]{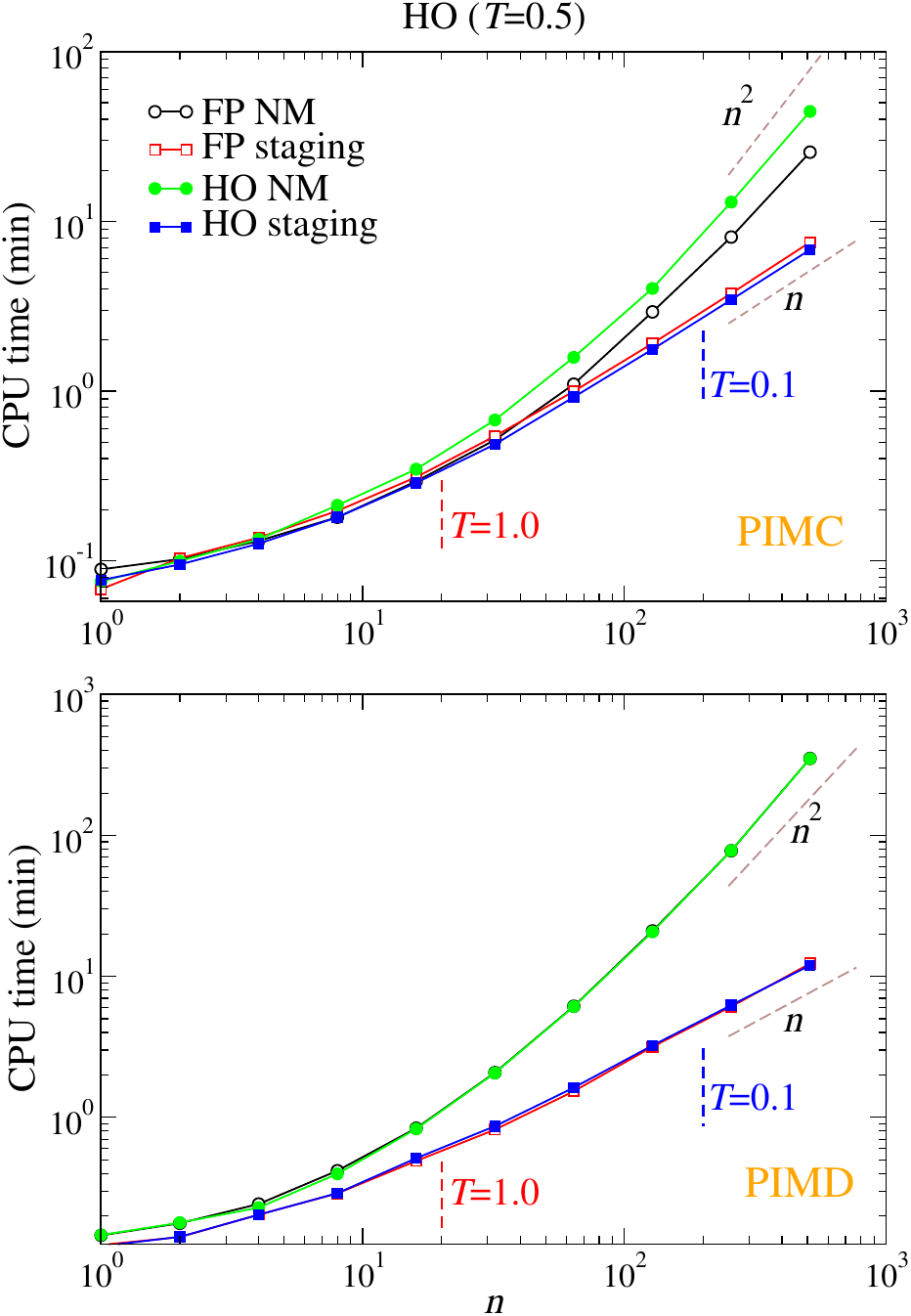}
\centering
\caption{System size dependence of CPU time for (simultaneously) computing $E$ and $C_{\rm V}$ of the AO model, at $T=0.5$. Both staging and NM coordinates are used with PIMC (top) and PIMD (bottom) simulations. The vertical dashed lines correspond to the minimum $n$ value ($20\beta\hbar\omega$ here) required for statistical convergence at the given $T$. } 
\label{fig:cpu} 
\end{figure}

Figure~\ref{fig:cpu} represents the computational cost in terms of the CPU time needed for (simultaneously) computing $E$ and $C_{\rm V}$ of the HO model as a function of  the number of beads. We consider PIMC (top) and PIMD (bottom) simulations in both HO and FP coordinates, at $T=0.5$ (similar at other temperatures). The vertical dashed lines correspond to the number of beads that would be used according to the $20\beta\hbar\omega$ criterion; i.e., $n=200$ and $20$ ate $T=0.1$ and $1.0$, respectively. We notice that, the computational expense in the HO and FP methods (staging or NM) is nearly similar. This is because the only difference between these methods is the transformation formula. At large $n$, results from both PIMC and PIMD simulations show the expected ${\cal O}\left(n\right)$ and ${\cal O}\left(n^2\right)$ asymptotic scalings associated with staging and NM coordinates, respectively. However, as mentioned earlier, the computational cost of the NM transformations could reduce to ${\cal O}\left(n\log\left(n\right)\right)$ using the FFT approach (not adopted here).

\section{CONCLUSIONS}
\label{sec:conclusion}
We have introduced novel staging coordinates that diagonalize the primitive path integral action of the harmonic oscillator (HO) model. These coordinates are exclusively applicable to systems with a harmonic character, such as some quantum oscillators and crystalline systems. In this context, the method can be viewed as a natural extension to the standard staging coordinates, which only diagonalize the kinetic contribution of the action, corresponding to the free particle (FP) model. Additionally, the FP based normal mode (NM) coordinates were trivially extended to diagonalize the entire HO action as well. In general, the staging coordinates are computationally preferable over the NM coordinates due its excellent ${\cal O}\left(n\right)$ CPU scaling with the number of beads, rather than ${\cal O}\left(n^2\right)$, or ${\cal O}\left(n\log\left(n\right)\right)$ using FFT, for the NM case. This becomes especially important at low temperatures, where large number of beads are required for statistical convergence. 

We provide implementation schemes for applying the HO staging and NM coordinates in both PIMC and PIMD simulation methods. As an application, we use a one-dimensional HO model to assess the sampling efficiency using the new HO staging and NM coordinates, in comparison to the respective FP-based coordinates. The sampling performance was represented in terms of the statistical uncertainty in the total energy $E$ and heat capacity at constant volume $C_{\rm V}$, using the centroid virial estimator. Moreover, the effect of anharmonicity on sampling was also investigated using a one-dimensional asymmetric anharmonic oscillator (AO). 

Results from both PIMC and PIMD methods consistently show higher precision when using the HO coordinates, in comparison to the FP coordinates, in the whole temperature range considered. Moreover, the  data suggest statistically indistinguishable performance from the staging and NM coordinates (for a given HO/FP method). This is to be expected because both coordinates diagonalize the same action. 

In PIMC simulations, uncertainties in both $E$ and $C_{\rm V}$ using the FP coordinates substantially increase (diverge) at low temperatures, whereas the HO coordinates show a smooth behavior at all temperature. This FP behavior is caused by the diverging nature of the PI ring-polymer size as the temperature decreases, which results in low acceptance rate and, hence, large uncertainty. Moreover, anharmonic effects did not show influence of the absolute uncertainties from all methods. 

For PIMD case, on the other hand, the sampling performance was investigated in terms of the maximum allowable time step size $\Delta t$, without loss of accuracy, and the statistical uncertainty (after using these time steps). The HO model was perfectly sampled (i.e., no dependence on $\Delta t$) using the HO coordinates. In this case, the allowed time step is only constraint by the maximum value for stable velocity Verlet integrator, which is given by $\Delta t=2/\omega$. However, much smaller time step sizes ($\approx 20-30\%$ less) were tolerated by the FP coordinates. For the AO model, the maximum $\Delta t$ limits decrease for all methods, yet the superior performance of the HO coordinates method, in terms of precision, is still preserved at all temperatures considered. While we adopted propagator splitting based on zero forces for the ``A'' step and a BOAOB scheme, we also considered other choices based on ring-polymer forces for ``A'' and OBABA ordering (see Secs. S6 and S7 of the Supporting Information). However, results indicate better performance with our choices for the PIMD integrator.

Moreover, we investigated the effect of anharmonicity in terms of the asymmetry of the AO model ($k_3$), using both PIMC and PIMD methods. We notice that the relative improvement in precision of the HO coordinates over the FP counterpart is preserved for the case of PIMD, irrespective to asymmetry. This indicates that the HO coordinates prescription for the fictitious masses (perfect for HO models) is still helpful even with anharmonicity. Of course, the relative performance depends on the nature of the actual system being studied (e.g., crystals). Similar conclusions can be made with the PIMC method at low-to-intermediate $k_3$ values. However, as $k_3$ approaches the unstable limit (second minimum is developed),  the relative improvement provided by the HO coordinates decreases.

Although we applied the HO staging and NM coordinates to only quantum oscillator models, the framework is readily applicable to quantum crystals, using both PIMC and PIMD methods. In such systems, the Einstein crystal contribution (self force constants of the Hessian) could act as independent harmonic oscillators, such that same analysis adopted here can be applied. Of course, these realistic models have inherent anharmonicity, such that detailed studies on the performance of HO coordinates are needed. We emphasize that the HO coordinates are \textit{not} applicable to fluids, or systems with no harmonic contribution (e.g., purely quartic models), or system with imaginary modes at the equilibrium configurations (both molecular bonds and crystals).

Although we only applied the HO coordinates to improve sampling, they can be used to modify the expressions of the PI estimators for an even more precise estimations. In fact, we are currently proceeding in this direction, using the harmonically mapped averaging (HMA) technique developed in our group.\cite{schultz2016reformulation} Moreover, while we only considered PI sampling approaches to determine static properties, extensions to dynamical properties (e.g., quantum time-correlations) could be possible, given that the system requirements mentioned above are met. An appealing feature of the HO coordinates if applied to approximate dynamical methods, such as ring-polymer molecular dynamics (RPMD) and centroid molecular dynamics (CMD), is that the centroid degree of freedom is inherently included in the coordinates, unlike the FP coordinates. With the simplicity of the HO coordinates and with no extra computational cost, these coordinates provide highly efficient alternative to traditional FP based coordinates when applied to systems with a harmonic character (either explicit or implicit) such as quantum molecular bonds and crystals.

\section{Supporting Information}
The Supporting Information document includes derivations of the equations used in the main text, along with finite-size effects results. We also present comparisons of the performance of our PIMD propagator (Eqs.~\ref{eq:pimd_splitting} and~\ref{eq:baoab}) to other schemes based on ring-polymer forces and OBABO ordering, using the FP staging and NM coordinates. The comparison also includes the recent (approximate) propagator based on the Cayley transformation.\cite{korol2019cayley}

\section*{ACKNOWLEDGMENTS}
Computational resources were provided by the Center for Computational Research (CCR), University at Buffalo, NY.

\bibliography{references}

\newpage
\begin{figure*} 
\centerline{\includegraphics[width=\textwidth]{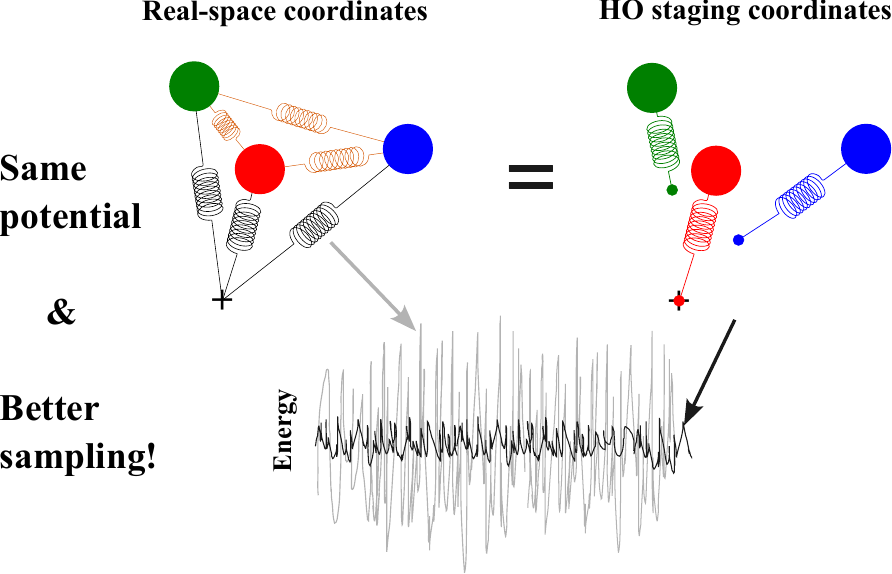}}
\caption{For Table of Contents Only}
\end{figure*}

\end{document}


\maketitle

\section{Derivation of $x_i^*$ Expression}
\label{sec:xistar_si}
\begin{figure}
\includegraphics[width=0.4\textwidth]{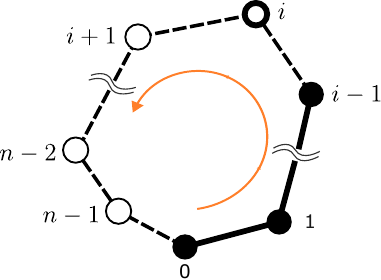}
\caption{Illustration of the counterclockwise process to determine the instantaneous equilibrium position $x_{i}^{*}$ of bead $i$, used with staging coordinates (Eq. 3 in the main text). This is accomplished by minimizing the effective potential $V$ with represent to the positions of beads $i$ through $n-1$ (open circles), while keeping the others fixed (filled circles). Lines represent PI harmonic interactions, with solid lines connecting the fixed beads, while dashed lines connecting the free beads. External interactions (not shown) are always assumed.}
\label{fig:staging_i}
\centering
\end{figure}

By definition (see Eqs. 3 and 4 of the main text), the $x_i^*$ position (for $i>0$) is obtained by minimizing $V$ with respect to $u_i$, while fixing all other staging coordinates,
\begin{align}
\label{eq:dVdui_1}
\frac{\partial V}{\partial u_i} = 0 
\rightarrow x_i^*, \;\; i>0
\end{align}
However, since $V\left({\bf x}\left({\bf u}\right),\beta\right)$ is a natural function in the Cartesian coordinates (Eq. 2 in the main text), we need to perform coordinate transformation in order to evaluate this derivative. Based on the definition of the HO staging transformation (Eq. 3 in the main text), the Cartesian coordinate of bead $i$ is given as $x_i\left(u_0,\dots,u_i\right)$. Using the chain rule, the derivative is then given as
\begin{align}
\label{eq:dVdui_2}
\frac{\partial V}{\partial u_i} = 
\sum_{j=i}^{n-1} \frac{\partial V\left({\bf x},\beta\right)}{\partial x_j'} 
\frac{\partial x_j'}{\partial u_i} = 0, \;\; i>0
\end{align}
where $j=i,i+1,\dots,n-1$.  we assume the beads $j=i$ to $j=n-1$ to be ``free'' for the purpose of derivatives, while the other beads ($0,\dots,i-1$) are at their actual locations. Because Eq.~\ref{eq:dVdui_2} holds for any configuration and $x_j'$ are independent variables, it is equivalent to the following set of conditions,
\begin{align}
\label{ed:dVdxjp}
\frac{\partial V\left({\bf x},\beta\right)}{\partial x_j'}=0,\;\; j=i,\dots,n-1
\end{align}
such that the $x_i'$ solution is $x_i^{*}$. Therefore, $x_i^*$ is defined as an instantaneous equilibrium position of bead $i$, such that $V$ is minimum with respect to hypothetically ``free'' beads of indices $i$ to $n-1$, while fixing the remaining beads at their actual locations.

Although we are interested in only the $x_i'$ solution, we have to (simultaneously) satisfy all the $n-i$ conditions given by Eq.~\ref{ed:dVdxjp}, for each bead $i$. Using the Cartesian space definition of $V$ (Eq. 1b in the main text), these conditions reduce to the following system of $n-i$ linear equations,
\begin{eqnarray}
\label{eq:Ax_stage}
\begin{pmatrix}
c & -1 &  & &  \\ 
-1 & c & -1 & & \\ 
  &  &  \ddots & & \\ 
  & & -1&  c& -1 \\ 
  & & & -1  & c 
\end{pmatrix}
\begin{pmatrix}
x'_i \\
x'_{i+1} \\
\vdots \\ 
x'_{n-2} \\
x'_{n-1} \\
\end{pmatrix}
= 
\begin{pmatrix}
x_{i-1} \\
 0 \\
\vdots \\ 
0 \\
x_{0} \\
\end{pmatrix}, \, i>0
\end{eqnarray}
where $c= 2 + \epsilon^2$, with $\epsilon\equiv\frac{\beta\hbar\omega}{n}=\frac{\omega}{\omega_n}$. This coefficients matrix is symmetric and tridiagonal, with constant coefficients along the diagonal (i.e., Toeplitz). Fortunately, the inverse of this matrix is known analytically, with the following $a,b$ components,\cite{hu1996}
\begin{align}
S_{a,b} &= \frac{\cosh\left(\left(n-i+1-|a-b|\right)\alpha\right)
-\cosh\left(\left(n-i+1-a-b\right)\alpha\right)}{2 \sinh\left(\alpha\right) \sinh\left(\left(n-i+1\right)\alpha\right)} 
\end{align}
where $a,b=1,\dots,n-i$ and $\alpha \equiv 2\sinh^{-1}\left(\frac{\epsilon}{2}\right)$. The $x'_i$ solution is then equal to $x_i^{*}\left(x_0,x_{i-1}\right) = S_{1,1} { x}_{i-1} + S_{1,n-i} { x}_{0}$, or
\begin{eqnarray}
\label{eq:xc_ho_staging_apdx}
x_i^{*}\left(x_0,x_{i-1}\right) =  
   \frac{\sinh\left(\alpha\right) { x}_0 + \sinh\left(\left(n-i\right)\alpha \right) { x}_{i-1}}{\sinh\left(\left(n-i+1\right) \alpha\right)},\, i>0
\end{eqnarray}
Accordingly, for our special PI harmonic action, $x_i^{*}$ depends solely on two positions, $x_0$ and $x_{i-1}$, of the fixed segment, which is a special case of the generalized definition (Eq. 3 in the main text). This is not surprising since the intermediate fixed beads ($1,\dots,i-2$) do not interact directly with bead $i$ (see Fig.~\ref{fig:staging_i}).

\section{Derivation of $k_i$ Expression}
\label{sec:ki_si}
Here we derive the expression for the HO staging spring constant $k_i$, through second derivatives of the PI effective potential of the HO model. Let us first express this potential in HO staging coordinates (Eq. 4 in the main text) in terms of the Cartesian coordinates using Eq. 3 of the main text,
\begin{align}
\label{eq:V_r_s_apdx}
V\left({\bf x},\beta\right) = \sum_{j=0}^{n-1} \frac{1}{2} \frac{m \omega_n^2}{n} \left(x_j -  x_{j-1} \right)^2
+ \sum_{j=0}^{n-1} \frac{1}{2} \frac{m \omega^2}{n} x_j^2 = \sum_{j=0}^{n-1} \frac{1}{2} k_j \left[x_j-x_j^{*}\left(x_0,x_{j-1}\right)\right]^2
\end{align}
We will start by deriving $k_i$ expression for $i>0$ and then the $i=0$ case. Taking the first derivative of both sides of Eq.~\ref{eq:V_r_s_apdx}, with respect to $x_i$ (for $i>0$), yields 
\begin{align}
 \frac{m \omega_n^2}{n} \left(2x_i-x_{i-1}-x_{i+1} \right)
+  \frac{m\omega^2}{n}  x_i 
=  k_i \left[x_i-x_i^{*}\left(x_0,x_{i-1}\right)\right] 
- k_{i+1} \left[x_{i+1}-x_{i+1}^{*}\left(x_0,x_i\right)\right]
\frac{\partial x_{i+1}^{*}\left(x_0,x_i\right)}{\partial x_i}, \; i>0
\end{align}
Then, taking the second derivative, and recognizing that the second derivative of $x_{i+1}^*$ with respect to $x_i$ is zero (see Eq.~\ref{eq:xc_ho_staging_apdx}), we get
\begin{equation}
k_i = 2 \frac{m \omega_n^2}{n}
+  \frac{m\omega^2}{n} 
- \left[\frac{\partial x_{i+1}^{*}\left(x_0,x_i\right)}{\partial x_i}\right]^2 k_{i+1} , \; i>0.
\end{equation}
According to Eq.~\ref{eq:xc_ho_staging_apdx}, the derivative of $x_i^*$ is
\begin{equation}
\frac{\partial x_{i+1}^{*}\left(x_0,x_i\right)}{\partial x_i}
= \frac{\sinh\left(\left(n-1-i\right)\alpha\right)}{\sinh\left(\left(n-i\right)\alpha\right)}, \; i>0.
\end{equation}
hence,
\begin{equation}
k_i = 2 \frac{m \omega_n^2}{n}
+  \frac{m\omega^2}{n} 
-  \frac{\sinh^2\left(\left(n-1-i\right)\alpha\right)}{\sinh^2\left(\left(n-i\right)\alpha\right)} k_{i+1}, \; i>0
\end{equation}
where, the spring constants $k_i$ can be obtained in a backward recursion manner as
\begin{subequations}
\begin{align}
k_{n-1} &= \frac{2 m \omega_n^2}{n}+\frac{m\omega^2}{n}  \nonumber\\
k_{n-2} &= \frac{2 m \omega_n^2}{n}+\frac{m\omega^2}{n}  - \frac{\sinh^2\left(\alpha\right)}{\sinh^2\left(2\alpha\right)} k_{n-1}\nonumber\\
k_{n-3} &= \frac{2 m \omega_n^2}{n}+\frac{m\omega^2}{n}  - \frac{\sinh^2\left(2\alpha\right)}{\sinh^2\left(3\alpha\right)} k_{n-2}\nonumber\\
\vdots \nonumber\\
k_{n-i} &= \frac{2 m \omega_n^2}{n}+\frac{m\omega^2}{n}  - \frac{\sinh^2\left(\left(n-1-i\right)\alpha\right)}{\sinh^2\left(\left(n-i\right)\alpha\right)} k_{n-i+1}\nonumber\\
\vdots \nonumber\\
k_1 &= \frac{2 m \omega_n^2}{n}+\frac{m\omega^2}{n}  - \frac{\sinh^2\left(\left(n-2\right)\alpha\right)}{\sinh^2\left(\left(n-1\right)\alpha\right)} k_2\nonumber
\end{align}
\end{subequations}
Eventually, evaluating these expressions yields a simple closed form,
\begin{equation}
\label{eq:ki_apdx}
k_i =  \frac{m \omega_n^2}{n}   \frac{\sinh\left(\left(n-i+1\right)\alpha\right)}{\sinh\left(\left(n-i\right)\alpha\right)}, \;i>0.    
\end{equation}

Same procedure is followed to get $k_0$. Taking the first derivative of both sides of Eq.~\ref{eq:V_r_s_apdx}, with respect to $x_0$, 
\begin{align}
\frac{m \omega_n^2}{n} \left(2x_0-x_{n-1}-x_{1} \right)
+  \frac{m\omega^2}{n}  x_0  
= k_0 x_0 - \sum_{i=1}^{n-1}  \left[x_{i}-x_{i}^{*}\left(x_0,x_{i-1}\right)\right]
\frac{\partial x_{i}^{*}\left(x_0,x_{i-1}\right)}{\partial x_0} k_i,
\end{align}
then, taking the second derivative, and rearrange, we get
\begin{equation}
k_0 = \frac{2 m \omega_n^2}{n} 
+  \frac{m\omega^2}{n}  
- \sum_{i=1}^{n-1} 
\left[\frac{\partial x_{i}^{*}\left(x_0,x_{i-1}\right)}{\partial x_0} \right]^2 k_i.
\end{equation}
According to Eq.~\ref{eq:xc_ho_staging_apdx}, the derivative of $x_i^*$ with respect to $x_0$ is
\begin{eqnarray}
\frac{\partial x_i^{*}\left(x_0,x_{i-1}\right)}{\partial x_0}
= \left\{
\begin{array}{ll}
       \frac{\sinh\left(\alpha\right)+ \sinh\left(\left(n-1\right)\alpha\right)}{\sinh\left(n\alpha\right)} & i=1 \\
\frac{\sinh\left(\alpha\right)}{\sinh\left(\left(n-i+1\right)\alpha\right)}, & i>1
\end{array} 
\right.     
\end{eqnarray}
and hence,
\begin{align}
k_0 = \frac{2 m \omega_n^2}{n} 
+  \frac{m\omega^2}{n}  
-  \left[ \frac{\sinh\left(\alpha\right)+ \sinh\left(\left(n-1\right)\alpha\right)}{\sinh\left(n\alpha\right)} \right]^2 k_1 
- \sum_{i=2}^{n-1}  \frac{\sinh^2\left(\alpha\right)}{\sinh^2\left(\left(n-i+1\right)\alpha\right)} k_i.
\end{align}
Plugging $k_i$ expressions (Eq.~\ref{eq:ki_apdx}) into $k_0$, and simplify, yields a simple closed form,
\begin{eqnarray}
k_0 = 2\frac{m \omega_n^2}{n} \sinh\left(\alpha\right) \tanh\left(\frac{n\alpha}{2}\right)    
\end{eqnarray}

\section{Verification of the HO Staging Formulation}
\label{si:verify_staging}
Here we verify that the staging coordinates expression of the HO effective energy (Eq. 4 in the main text) is equivalent to the Cartesian coordinates one (Eq. 2 in the main text). Let us first plug the definition of $x_i^*$ (Eq. 4) into the HO effective potential form (Eq. 4 in the main text), such that it is only a function of the Cartesian coordinates,
\begin{align}
\label{eq:V_HO_stage_apdx}
V\left({\bf x},\beta\right) =  \frac{1}{2} k_0 x_0^2 
    + \sum_{i=1}^{n-1} \frac{1}{2} k_i 
    \left[x_i - \frac{\sinh\left(\alpha\right) { x}_0 + \sinh\left(\left(n-i\right)\alpha \right) { x}_{i-1}}{\sinh\left(\left(n-i+1\right) \alpha\right)}\right]^2.
\end{align}
The task now is to show that this expression is equivalent to the original potential in Cartesian coordinates (Eq. 2 in the main text). We achieve this by showing that both expressions have the same minimum (first derivative) and curvature (second derivative). Since both expressions are quadratic, higher order derivatives are zero in both cases. Therefore, same first and second derivatives is equivalent to same potential functions.

The configurations correspond to the minimum of the potential in  Cartesian coordinates (Eq. 2 in the main text) are $x_i=0$, with a zero potential. For the staging coordinates, the first derivative of Eq.~\ref{eq:V_HO_stage_apdx}, for $i>0$, is
\begin{align}
\label{eq:dV_ig0_apdx}
0=\frac{\partial V\left({\bf x},\beta\right)}{\partial x_i}
&= 
k_i \left[x_i - \frac{\sinh\left(\alpha\right) { x}_0 + \sinh\left(\left(n-i\right)\alpha \right) { x}_{i-1}}{\sinh\left(\left(n-i+1\right) \alpha\right)}\right]\nonumber\\
&- k_{i+1} \frac{\sinh\left(\left(n-i-1\right)\alpha \right)}{\sinh\left(\left(n-i\right) \alpha\right)} 
\times
\left[x_{i+1} - \frac{\sinh\left(\alpha\right) { x}_0 + \sinh\left(\left(n-i-1\right)\alpha \right) { x}_{i}}{\sinh\left(\left(n-i\right) \alpha\right)}\right], i>0.
\end{align}
Similarly, the $i=0$ case is
\begin{align}
\label{eq:dV_i0_apdx}
0=\frac{\partial V\left({\bf x},\beta\right)}{\partial x_0} 
&= 
k_0 x_0 - k_1 \frac{\sinh\left( \alpha\right) + \sinh\left( \left(n-1\right)\alpha\right)}{\sinh\left(n\alpha\right)} 
 \left[x_1 - \frac{\sinh\left(\alpha\right) + \sinh\left(\left(n-1\right)\alpha \right) }{\sinh\left(n \alpha\right)} x_0 \right] \nonumber\\
&- \sum_{j=2}^{n-1} k_j \frac{\sinh\left( \alpha\right)}{\sinh\left(\left(n-j+1\right) \alpha\right)} \left[x_j - \frac{\sinh\left(\alpha\right) { x}_0 + \sinh\left(\left(n-j\right)\alpha \right) { x}_{j-1}}{\sinh\left(\left(n-j+1\right) \alpha\right)}\right] \nonumber\\
\end{align}
Since Eqs.~\ref{eq:dV_ig0_apdx} and~\ref{eq:dV_i0_apdx} are valid for any $x_i$, these equations are satisfied if and only if all $x_i=0$, which is the same minimum of the  Cartesian space formulation.

Next, the second derivative of the Cartesian space potentials (Eq. 2 in the main text) with respect to $x_i$ is $2\frac{m\omega_n^2}{n} + \frac{m\omega^2}{n}$. For staging, using Eq.~\ref{eq:dV_ig0_apdx}, the second derivative for the $i>0$ case is 
\begin{align}
&\frac{\partial^2 V\left({\bf x},\beta\right)}{\partial x_i^2} 
= 
k_i + k_{i+1} \frac{\sinh^2\left(\left(n-i-1\right)\alpha \right) { x}_{i}}{\sinh^2\left(\left(n-i\right) \alpha\right)} \nonumber\\
&= \frac{m \omega_n^2}{n}  \left[ \frac{\sinh\left(\left(n-i+1\right)\alpha\right)}{\sinh\left(\left(n-i\right)\alpha\right)}
+ \frac{\sinh\left(\left(n-i-1\right)\alpha \right) { x}_{i}}{\sinh\left(\left(n-i\right) \alpha\right)}
\right] \nonumber\\
&= 2\frac{m \omega_n^2}{n} \cosh\left(\alpha\right) \nonumber\\
&= 2\frac{m \omega_n^2}{n} + \frac{m\omega^2}{n},   i>0
\end{align}
which matches the second derivative of the Cartesian coordinates. For the $i=0$ case, taking the first derivative of Eq.~\ref{eq:dV_i0_apdx} with respect to $x_i$ yields
\begin{align}
&\frac{\partial^2 V\left({\bf x},\beta\right)}{\partial x_0^2} = 
k_0 
+ k_1 
\left[\frac{\sinh\left(\alpha\right) + \sinh\left(\left(n-1\right)\alpha \right) }{\sinh\left(n \alpha\right)}\right]^2 \nonumber\\
&+\sum_{j=1}^{n-1} k_j \left[\frac{\sinh\left( \alpha\right)}{\sinh\left(\left(n-j+1\right) \alpha\right)} \right]^2 \nonumber\\
&= 
2\frac{m \omega_n^2}{n} \sinh\left(\alpha\right) \tanh\left(\frac{n\alpha}{2}\right)  \nonumber\\
&+ \frac{m\omega_n^2}{n} \frac{\left(\sinh\left(\alpha\right) + \sinh\left(\left(n-1\right)\alpha \right) \right)^2}{\sinh\left(\left(n-1\right) \alpha\right)\sinh\left(n \alpha\right)} \nonumber\\
&+ \frac{m\omega_n^2}{n} \sum_{j=2}^{n-1}  \frac{\sinh^2\left( \alpha\right)}{\sinh\left(\left(n-j\right)\alpha\right) \sinh\left(\left(n-j+1\right) \alpha\right)} \nonumber\\
&= 2\frac{m\omega_n^2}{n} + \frac{m\omega^2}{n},
\end{align}
where the last line is obtained from Mathematica.\cite{mathematica} This completes the verification that the staging expression (Eq.~\ref{eq:V_HO_stage_apdx}) is equal to the  Cartesian space one (Eq. 2 in the main text).

\section{Clockwise Version of Staging}
\label{si:clockwise_staging}
The staging expressions derived in this work are based on a forward ($0,1,\dots,n-1$) counterclockwise representation (Fig.~\ref{fig:staging_i}). However, the traditional FP staging known in literature is based on a backward ($0, n-1,n-2,\dots,1$) clockwise formulation. Clearly, there is no difference between both versions since only beads labelling is different. However, to show equivalence between both FP staging versions and to provide an alternative option for HO staging, which might be of more convenience to some, we present here the clockwise version of both HO and FP stagings. This achieved by simply replacing the $x_{i-1}$ position by $x_{i+1}$ and the $n-i$ index by $i$ (not including positions subscripts), which results in the following HO and FP staging expressions.

\subsection*{Clockwise HO Staging}
The clockwise version of the HO staging coordinates (Eq. 7 in the main text) is given by a non-recursive form,
\begin{eqnarray}
\label{eq:r2x_transformation_clockwise}
u_i = x_i - \frac{\sinh\left(\alpha\right) x_0 + \sinh\left(i \alpha \right) x_{i+1} }{\sinh\left(\left(i+1\right) \alpha\right)}, & i>0
\end{eqnarray}
where $u_0=x_0$ and $x_n=x_{0}$. Accordingly the inverse transformation (${\bf x}\leftarrow {\bf u}$) is given by the following recursive (from $i=n-1$ to $1$) relation:
\begin{eqnarray}
\label{eq:u2x_transformation_clockwise}
x_i = u_i + \frac{\sinh\left(\alpha\right) u_0 + \sinh\left(i \alpha \right) x_{i+1} }{\sinh\left(\left(i+1\right) \alpha\right)}, & i>0
\end{eqnarray}
Similarly, the clockwise version of the staging spring constants (Eq. 6 in the main text) are  
\begin{eqnarray}
\label{eq:ki_HO_staging_clockwise}
k_i = \frac{m\omega_n^2}{n}\left\{
\begin{array}{ll}
      2 \sinh\left(\alpha\right) \tanh\left(\frac{n\alpha}{2}\right), & i=0 \\
       \frac{\sinh\left(\left(i+1\right)\alpha\right)}{\sinh\left(i\alpha\right)}, & i>0
\end{array} 
\right. 
\end{eqnarray}

\subsection*{Clockwise FP Staging}
For FP staging, the clockwise version of the staging transformation (Eq.9  in the main text) is given by
\begin{eqnarray}
\label{eq:r2x_transformation_clockwise}
u_i = x_i - \frac{ x_0 +  i x_{i+1}}{i+1}, & i>0
\end{eqnarray}
where $u_0=x_0$ and $x_n=x_{0}$. From this, the inverse transformation (${\bf x}\leftarrow {\bf u}$) is given by the following  recursive (from $i=n-1$ to $1$) relation:
\begin{eqnarray}
\label{eq:u2x_transformation_clockwise}
x_i = u_i + \frac{ u_0 +  i  x_{i+1}}{i+1}, & i>0
\end{eqnarray}
Similarly, the clockwise version of the FP staging spring constant is given from Eq. 11 of the main text as,
\begin{eqnarray}
\label{eq:ki_HO_staging_clockwise}
k_i =  \left\{
\begin{array}{ll}
     0, & i=0 \\
       \frac{m \omega_n^2}{n} \frac{i+1}{i}, & i>0
\end{array} 
\right. 
\end{eqnarray}

\begin{figure}
\includegraphics[width=\textwidth]{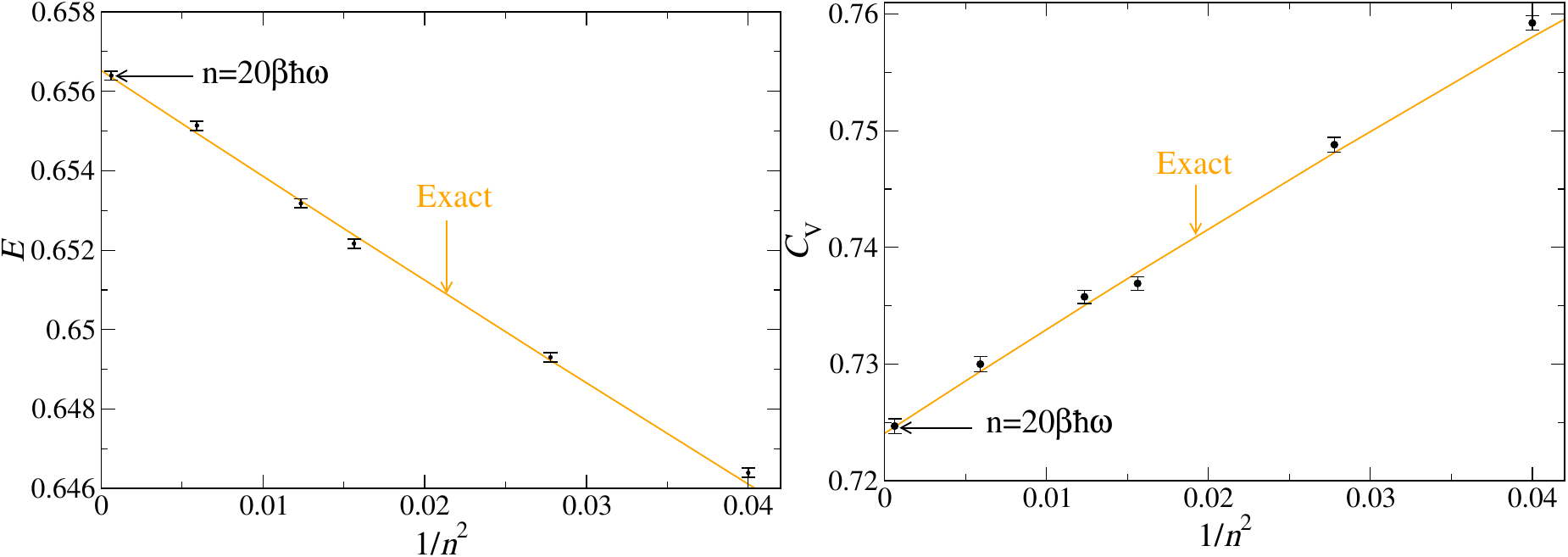}
\centering
\caption{Dependence of the energy (top) and heat capacity (bottom) on the number of beads for the HO model, at $T=0.5$. The orange lines represent the exact values (Eq. 39 in the main text). The leftmost point ($n=40$) corresponds to our $20\beta\hbar\omega$ criterion for statistical convergence. The data were generated using HO staging coordinates in PIMC.}
\label{fig:conv_n}
\end{figure}

\section{Finite Size Effects}
\label{si:fse}
Figure~\ref{fig:conv_n} shows the finite size effects (FSE) for the energy (top) and heat capacity (bottom) of the HO model, at $T=0.5$. We choose the HO model in order to verify the accuracy of our estimations against the exact analytical values of the model. However, similar results are observed with the AO model (not shown). Since the leading FSE term is known to be $\left(\beta\hbar\omega/n\right)^2$,\cite{} we represent the data in terms of $1/n^2$. Within the number of beads range considered ($n=5$ to $40$), it is evident that the trend of both $E$ and $C_{\rm V}$ is linear, which is consistent with the leading term behavior. The orange lines represent the analytical values (Eq. 39 in the main text), which clearly show the accuracy of our estimates at all system sizes considered. 

\begin{figure}
\includegraphics[width=\textwidth]{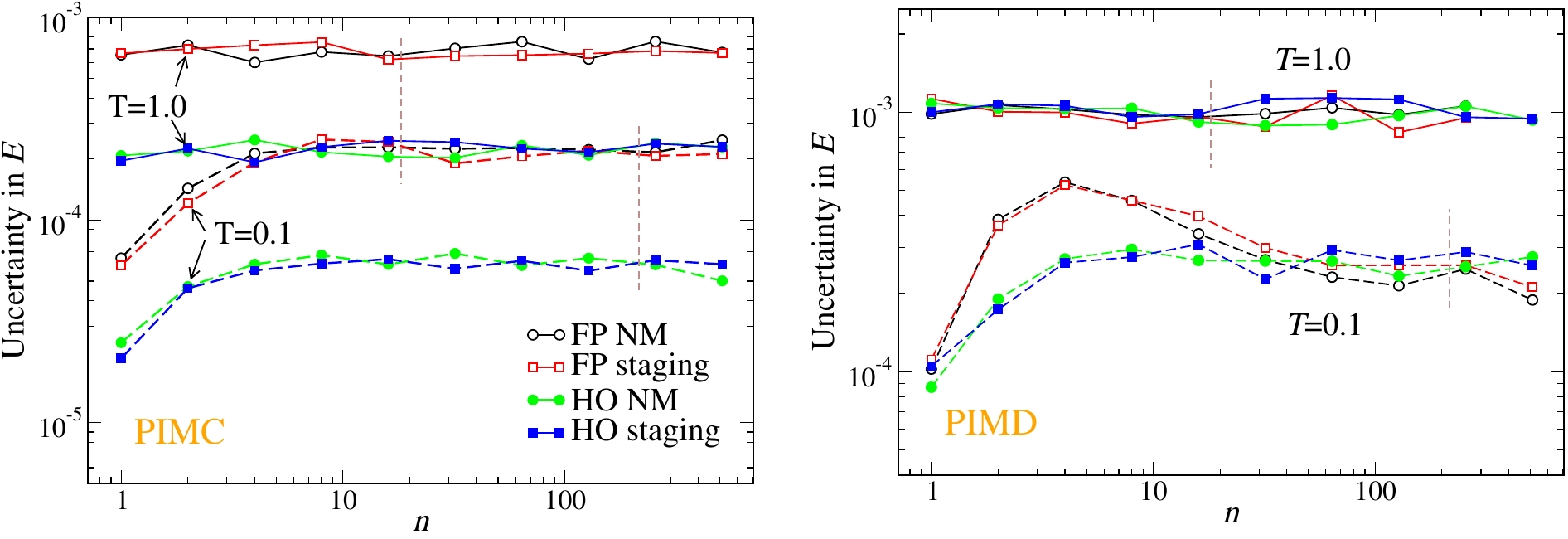}
\centering
\caption{Dependence of the uncertainty in energy on the number of beads, $n$, for the HO model, using PIMC (top) and PIMD (bottom) at low ($T=0.1$) and high ($T=1.0$) temperatures. The vertical lines correspond to the $n=20\beta\hbar\omega$ criterion for statistical convergence; see Fig.~\ref{fig:conv_n}.}
\label{fig:err_n}
\end{figure}

Moreover, as mentioned earlier, we use number of beads equal to $n=20\beta\hbar\omega$, to ensure converged results. To verify this, we consider this specific value in Fig.~\ref{fig:conv_n} (leftmost data point). It is clear that the energy and heat capacity data at that point are statistically consistent with the continuum limit (Eq. 40 in the main text). Similar observations were reported (not shown) using other temperatures and HO frequencies, for both HO and AO models.

Figure~\ref{fig:err_n} shows the effect of system size on the precision of estimating the energy of HO model, using PIMC (top) and PIMD (bottom) simulations. We present both HO and FP based coordinates, at low ($T=0.1$, dashed lines) and high ($T=1.0$, solid lines) temperatures. We do not find a noticeable difference in the behavior between the HO and FP methods. In addition, the  behavior beyond the converged number of beads ($n=20\beta\hbar\omega$, represented by vertical lines) is flat, which is a well established observation in the literature.

\section{Alternative PIMD Splitting for the FP Coordinates}
\label{sec:exactA_si}
For the FP coordinates (both staging and NM), a common alternative to the PIMD splitting we have in the main text (Eq. 21) is to decompose the total force $F_l$ into a ring polymer ($F_l^{\rm RP}$) and a physical ($F_l^{\rm phy}$) contributions.\cite{ceriotti2010efficient,liu2016simple} Then, associate the former and the latter to the ``A'' and ``B'' parts of the propagator, respectively, such that the equation of motion becomes
\begin{eqnarray}
\label{eq:conv_splitting_si}
\begin{pmatrix}
 \dot{q}_l \\
 \dot{p}_l
\end{pmatrix}
=
\underbrace{\begin{pmatrix}
 \frac{p_l}{m_l} \\
F_l^{\rm RP}
\end{pmatrix}}_\text{A}
+
\underbrace{\begin{pmatrix}
 0 \\
F_l^{\rm phy}
\end{pmatrix}}_\text{B}
+ 
\underbrace{\begin{pmatrix}
 0 \\
-\gamma p_l + \sqrt{2m_lk_{\rm B}T\gamma} \; \xi_l
\end{pmatrix}}_\text{O},
\end{eqnarray}
where $F_l^{\rm RP}=-m_l \Omega_n^2  q_l$ for the non-zero modes and $F_{0}^{\rm RP}=0$ for the zero-mode, and $F_l^{\rm phy} = -\frac{\partial U}{\partial q_l} $, with $U\equiv \frac{1}{n} \sum_{i=0}^{n-1}   U\left({\bf x}_i\right)$. Here, $\Omega_n^2\equiv \frac{\omega_n^2}{n}$ represents a common frequency for all the non-zero modes, with $\omega_n=n/\beta\hbar$ as defined in the main text. The motivation for this formulation is that the ``A'' contribution represents the standard harmonic oscillator propagator, 
\begin{align}
\label{eq:harm_eom_si}
\begin{pmatrix}
 \dot{q}_l \\
 \dot{p}_l
\end{pmatrix} 
=  A \begin{pmatrix}
 {q}_l \\
 {p}_l
\end{pmatrix}, \; \text{where} \;
A  = \begin{pmatrix}
0 & \frac{1}{m_l} \\
-m_l \Omega_n^2 & 0
\end{pmatrix}
\end{align}
which has an \textit{exact} solution at an arbitrary time,
\begin{align}
\begin{pmatrix}
 {q}_l\left(t\right) \\
 {p}_l\left(t\right)
\end{pmatrix} 
=  \exp{\left(A\Delta t\right)} 
\begin{pmatrix}
 {q}_l\left(0\right) \\
 {p}_l \left(0\right)
\end{pmatrix} \nonumber = 
\begin{pmatrix}
\cos\left(\Omega_n  t\right) & \frac{\sin\left(\Omega_n  t\right)}{m_l\Omega_n} \\
-m_l \Omega_n \sin\left(\Omega_n  t\right) & \cos\left(\Omega_n  t\right)
\end{pmatrix} 
\begin{pmatrix}
 {q}_l\left(0\right) \\
 {p}_l \left(0\right)
\end{pmatrix}, l>0
\end{align}
The zero-mode has a zero force and, hence, it represents a free-flight motion, $q_0(t)=q_0(0)+\frac{p_0(0)}{m_0} t$, where $m_0=m$ (Eq. 32 in the main text). We present below two implementations for this splitting, namely: exact and approximate ``A'' propagators.

\subsection{Exact ``A'' Propagator}
Using the exact solution of the ``A'' propagator in Eq.~\ref{eq:conv_splitting_si}, the BAOAB procedure is given by 
\begin{subequations}
\label{eq:baoab_exactA_si}
\begin{align}
{\rm B}\left(\frac{\Delta t}{2}\right): \;  & p_l \leftarrow p_l  + F_l^{\rm phy} \frac{\Delta t}{2}  \\
{\rm A} \left(\frac{\Delta t}{2}\right): \; & 
\left\{\begin{array}{l}
q_0 \leftarrow  q_0 +\frac{p_0}{m_0} \frac{\Delta t}{2},  \; l=0,\\
\begin{pmatrix}
 {q}_l \\
 {p}_l
\end{pmatrix} 
\leftarrow 
\begin{pmatrix}
\cos\left(\Omega_n  \frac{\Delta t}{2}\right) & \frac{\sin\left(\Omega_n  \frac{\Delta t}{2}\right)}{m_l\Omega_n} \\
-m_l \Omega_n \sin\left(\Omega_n  \frac{\Delta t}{2}\right) & \cos\left(\Omega_n  \frac{\Delta t}{2}\right)
\end{pmatrix} 
\begin{pmatrix}
 {q}_l \\
 {p}_l
\end{pmatrix}, l>0 
\end{array}\right.
\\
{\rm O} \left(\Delta t\right): \; &p_l \leftarrow p_l \exp\left(-\gamma \Delta t\right) +  \sqrt{m_l k_{\rm B}T\left(1-\exp\left(2\gamma\Delta t\right)\right)} \; R_l\\
{\rm A} \left(\frac{\Delta t}{2}\right): \; & 
\left\{\begin{array}{l}
q_0 \leftarrow  q_0 +\frac{p_0}{m_0} \frac{\Delta t}{2},  \; l=0,\\
\begin{pmatrix}
 {q}_l \\
 {p}_l
\end{pmatrix} 
\leftarrow 
\begin{pmatrix}
\cos\left(\Omega_n  \frac{\Delta t}{2}\right) & \frac{\sin\left(\Omega_n  \frac{\Delta t}{2}\right)}{m_l\Omega_n} \\
-m_l \Omega_n \sin\left(\Omega_n  \frac{\Delta t}{2}\right) & \cos\left(\Omega_n  \frac{\Delta t}{2}\right)
\end{pmatrix} 
\begin{pmatrix}
 {q}_l \\
 {p}_l
\end{pmatrix}, l>0 
\end{array}\right.
\\
{\rm B} \left(\frac{\Delta t}{2}\right): \; &p_l \leftarrow  p_l + F_l^{\rm phy} \frac{\Delta t}{2}
\end{align}
\end{subequations}
where the $m_i$ masses in these FP coordinates are given in the main text (Eqs.32 for the FP staging and Eqs. 33 for the FP NM).

\subsection{Approximate ``A'' Propagator: Cayley Transformation }
\label{sec:www3_si}
In order to enhance the propagator stability, Korol \textit{et al.} introduced an \textit{approximate} propagator for the ``A'' part of Eq.~\ref{eq:conv_splitting_si}.\cite{korol2019cayley} In this formulation, the ``A'' propagator is approximated using the Cayley transformation as follows, 
\begin{align}
\label{eq:cayleyA_def}
\exp{\left(A\Delta t\right)} \approx {\rm cay}\left(A \Delta t\right)    &\equiv \left(I-\frac{\Delta t}{2} A\right)^{-1} \left(I+\frac{\Delta t}{2} A\right) 
\end{align}
where $I$ is a $2\times 2$ unity matrix. Using the definition of matrix $A$ (Eq.~\ref{eq:harm_eom_si}) yields,
\begin{align}
\label{eq:cayleyA}
{\rm cay}\left(A \Delta t\right)   
&=\begin{pmatrix}
1 & -\frac{\Delta t}{2m_l} \\
\frac{m_l \Omega_n^2\Delta t}{2} & 1
\end{pmatrix}^{-1}
\begin{pmatrix}
1 & \frac{\Delta t}{2m_l} \\
-\frac{m_l \Omega_n^2\Delta t}{2} & 1
\end{pmatrix}\nonumber\\
&=\frac{1}{1+\left(\frac{\Omega_n\Delta t}{2}\right)^2}\begin{pmatrix}
1 & \frac{\Delta t}{2m_l} \\
-\frac{m_l \Omega_n^2\Delta t}{2} & 1
\end{pmatrix}
\begin{pmatrix}
1 & \frac{\Delta t}{2m_l} \\
-\frac{m_l \Omega_n^2\Delta t}{2} & 1 
\end{pmatrix}
\nonumber\\
&=\frac{1}{1+\left(\frac{\Omega_n\Delta t}{2}\right)^2}\begin{pmatrix}
1-\left(\frac{\Omega_n\Delta t}{2}\right)^2 & \frac{\Delta t}{m_l} \\
- m_l \Omega_n^2\Delta t & 1-\left(\frac{\Omega_n\Delta t}{2}\right)^2
\end{pmatrix}
\end{align}
This propagator is symplectic, time-reversible, and has a unity Jacobian (unitary), and satisfies the strong stability criterion.\cite{korol2019cayley}

Using this ``A'' propagator, with a half time step, results in the following BAOAB procedure,
\begin{subequations}
\label{eq:BAOAB_cayley_si}
\begin{align}
{\rm B}\left(\frac{\Delta t}{2}\right): \;  & p_l \leftarrow p_l  + F_l^{\rm phy} \frac{\Delta t}{2m_l}  \\
{\rm A} \left(\frac{\Delta t}{2}\right): \; & 
\left\{\begin{array}{l}
q_0 \leftarrow  q_0 +\frac{p_0}{m_0} \frac{\Delta t}{2m_l},  \; l=0,\\
\begin{pmatrix}
 {q}_l \\
 {p}_l
\end{pmatrix} 
\leftarrow 
\frac{1}{1+\left(\frac{\Omega_n\Delta t}{4}\right)^2}\begin{pmatrix}
1-\left(\frac{\Omega_n\Delta t}{4}\right)^2 & \frac{\Delta t}{2m_l} \\
- m_l \Omega_n^2\frac{\Delta t}{2} & 1-\left(\frac{\Omega_n\Delta t}{4}\right)^2
\end{pmatrix}
\begin{pmatrix}
 {q}_l \\
 {p}_l
\end{pmatrix}, l>0 
\end{array}\right.
\\
{\rm O} \left(\Delta t\right): \; &p_l \leftarrow p_l \exp\left(-\gamma \Delta t\right) +  \sqrt{m_l k_{\rm B}T\left(1-\exp\left(2\gamma\Delta t\right)\right)} \; R_l\\
{\rm A} \left(\frac{\Delta t}{2}\right): \; & 
\left\{\begin{array}{l}
q_0 \leftarrow  q_0 +\frac{p_0}{m_0} \frac{\Delta t}{2m_l},  \; l=0,\\
\begin{pmatrix}
 {q}_l \\
 {p}_l
\end{pmatrix} 
\leftarrow 
\frac{1}{1+\left(\frac{\Omega_n\Delta t}{4}\right)^2}\begin{pmatrix}
1-\left(\frac{\Omega_n\Delta t}{4}\right)^2 & \frac{\Delta t}{2m_l} \\
- m_l \Omega_n^2\frac{\Delta t}{2} & 1-\left(\frac{\Omega_n\Delta t}{4}\right)^2
\end{pmatrix}
\begin{pmatrix}
 {q}_l \\
 {p}_l
\end{pmatrix}, l>0 
\end{array}\right.
\\
{\rm B} \left(\frac{\Delta t}{2}\right): \; &p_l \leftarrow  p_l + F_l^{\rm phy} \frac{\Delta t}{2m_l}
\end{align}
\end{subequations}
Note that, for the ``A'' part, we replaced $\Delta t$ in Eq.~\ref{eq:cayleyA} by $\Delta t/2$.

\subsection{Results}
\begin{figure}
\includegraphics[width=0.77\textwidth]{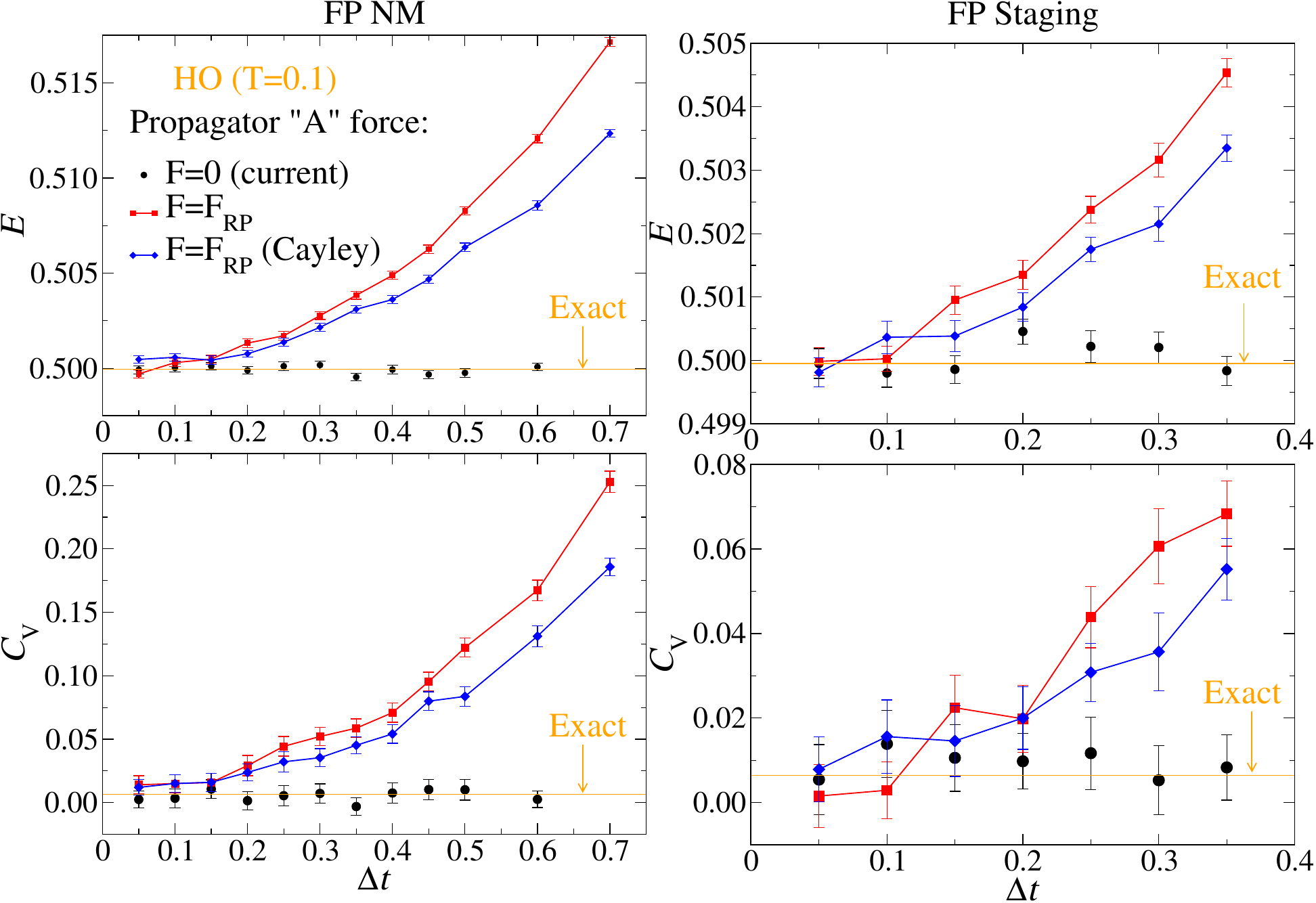}
\caption{Effect of the ``A'' propagator choice on the convergence of the total energy (top) and heat capacity (bottom) of the HO at $T=0.1$ with respect to the PIMD time step size, using both the FP NM (left) and staging (right) coordinates. Three choices for the forces of the ``A'' part are considered: $F=0$ (Eq. 21 in the main text) and using the ring-polymer intramolecular forces (Eq.~\ref{eq:conv_splitting_si}), with the exact (Eq.~\ref{eq:baoab_exactA_si}) and approximate (Cayley; Eq.~\ref{eq:BAOAB_cayley_si}) propagators. The exact results for the HO and AO are given using Eq. 39 of the main text and the NMM method, respectively.}
\label{fig:ECV_dt_A_ho_si}
\centering
\end{figure}

\begin{figure}
\includegraphics[width=0.77\textwidth]{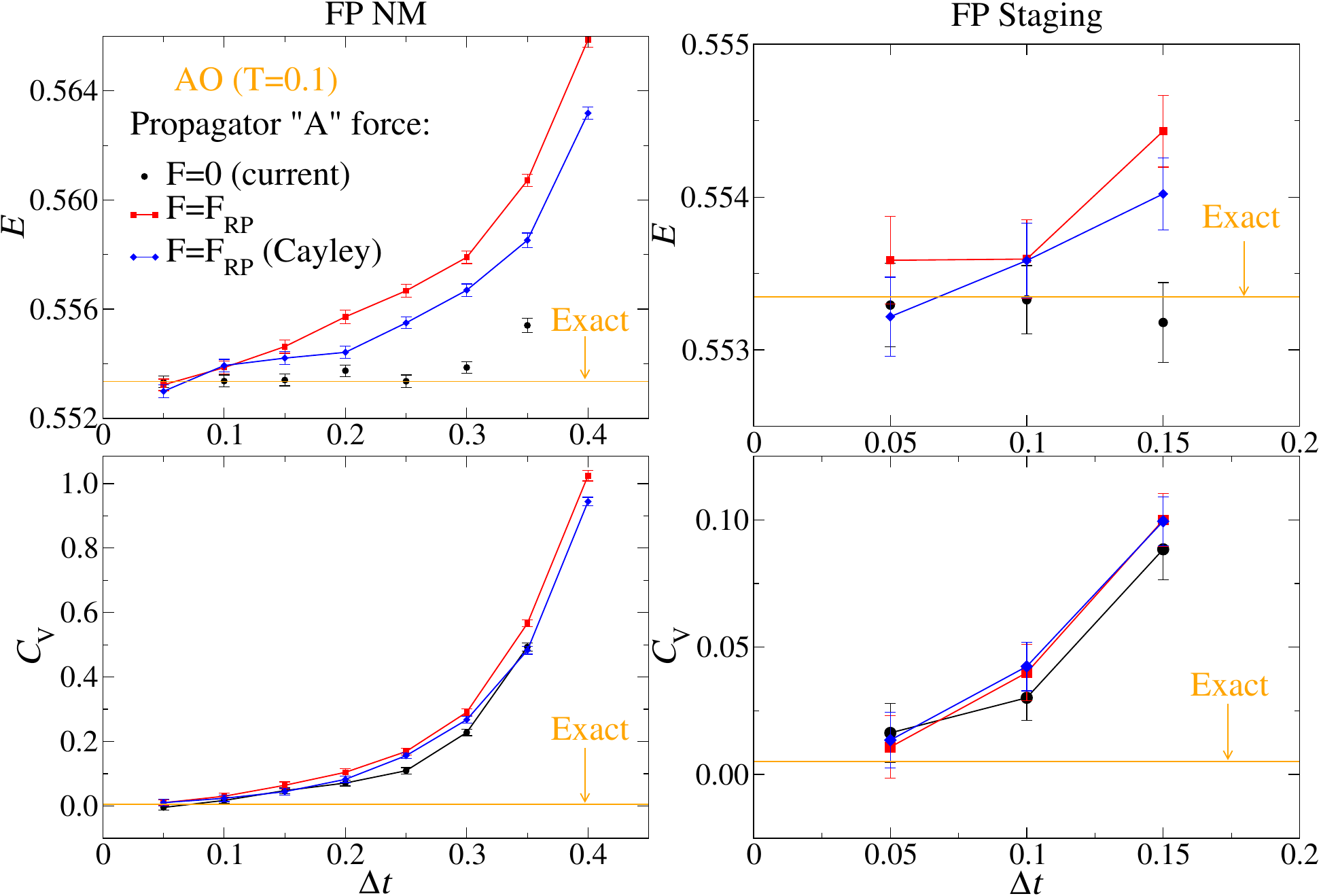}
\caption{Same as Fig.~\ref{fig:ECV_dt_A_ho_si}, but for the AO model.}
\label{fig:ECV_dt_A_ao_si}
\centering
\end{figure}

Figures~\ref{fig:ECV_dt_A_ho_si} and~\ref{fig:ECV_dt_A_ao_si} show the effect of different choices of the ``A'' propagator on the convergence of the energy and heat capacity of the HO and AO models, respectively, with respect to time step size. For the HO model, the original splitting we have in the main text for ``A'' part ($F=0$, Eq. 21) shows a superior performance over the ring-polymer splitting (Eq.~\ref{eq:conv_splitting_si}), using both the exact (~\ref{eq:baoab_exactA_si}) and approximate (Eq.~\ref{eq:BAOAB_cayley_si}) propagators. For the AO mode, on the other hand, Fig.~\ref{fig:ECV_dt_A_ao_si} shows that the original splitting provides a better convergence for the energy, relative to the ring-polymer splitting, while no improvement is introduced for the heat capacity case. Therefore, these results justify our choice for the ``A'' propagator we have in the main text (Eq. 21).

\newpage
\section{OBABO Propagator for the FP Coordinates}
\label{sec:OBABO_si}
Here, we consider the OBABO splitting order, which is another commonly used approach for PIMD simulations.\cite{ceriotti2010efficient,korol2019cayley,liu2016simple} Given that the performance of the current choice of the ``A'' part ($F=0$; Eq. 21 in the main text) yields a better performance as shown in the previous section, we consider this splitting choice in the OBABO ordering. The corresponding OBABO splitting procedure is then 
\begin{subequations}
\label{eq:ababo_si}
\begin{align}
{\rm O} \left(\frac{\Delta t}{2}\right): \; &p_l \leftarrow p_l \exp\left(-\gamma \frac{\Delta t}{2}\right) +  \sqrt{m_l k_{\rm B}T\left(1-\exp\left(\gamma\Delta t\right)\right)} \; R_l\\
{\rm B}\left(\frac{\Delta t}{2}\right): \;  &p_l \leftarrow p_l  + F_l \frac{\Delta t}{2}  \\
{\rm A} \left(\Delta t\right): \; &q_l \leftarrow q_l  + \frac{p_l}{m_l} \Delta t\\
{\rm B} \left(\frac{\Delta t}{2}\right): \; &p_l \leftarrow  p_l + F_l \frac{\Delta t}{2}\\
{\rm O} \left(\frac{\Delta t}{2}\right): \; &p_l \leftarrow p_l \exp\left(-\gamma \frac{\Delta t}{2}\right) +  \sqrt{m_l k_{\rm B}T\left(1-\exp\left(\gamma\Delta t\right)\right)} \; R_l
\end{align}
\end{subequations}

\subsection{Results}
\begin{figure}
\includegraphics[width=0.8\textwidth]{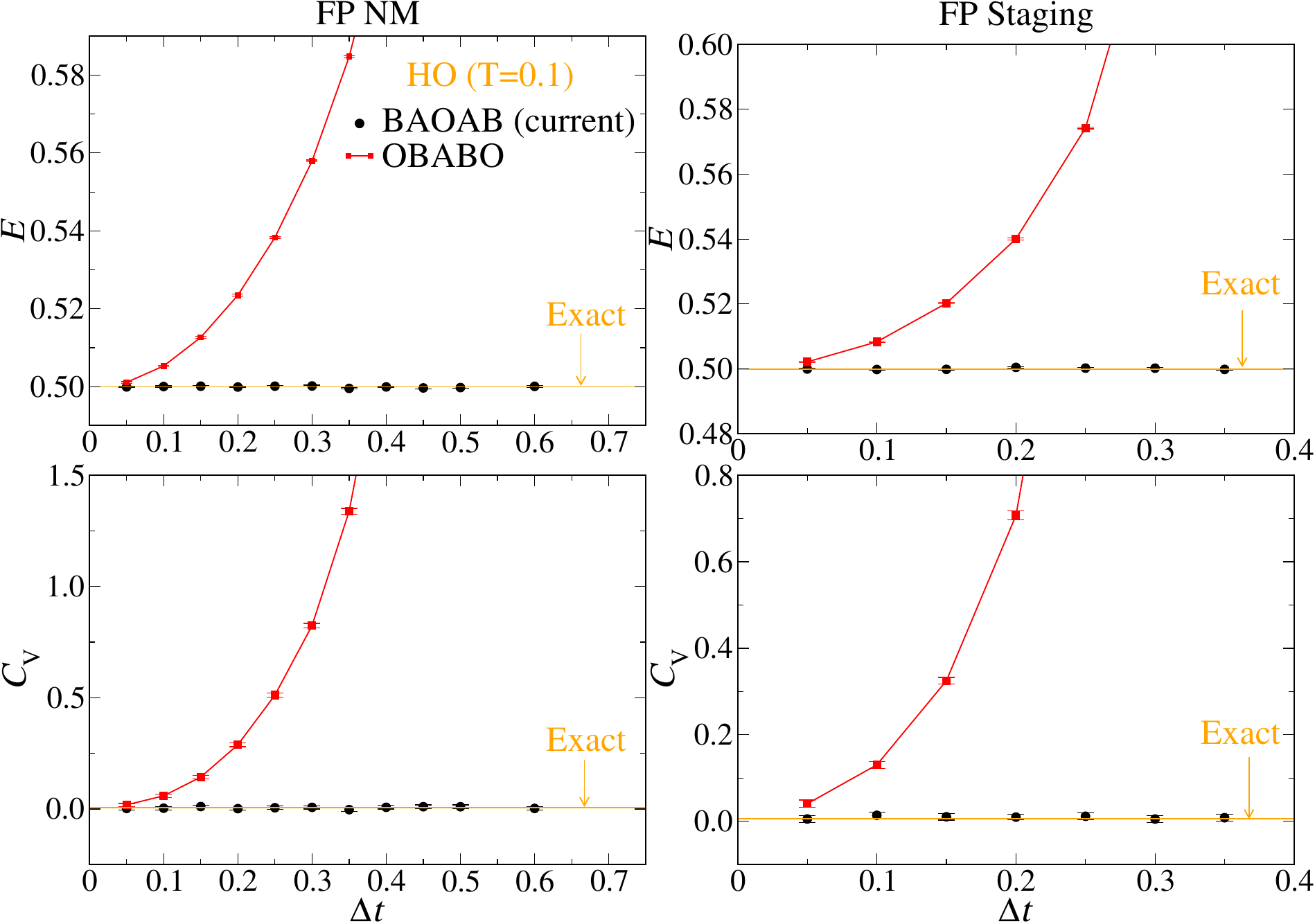}
\caption{Effect of the PIMD splitting choices, BAOAB (black) and OBABO (red), on the convergence of the energy (top) and heat capacity (bottom) of the HO at $T=0.1$ with respect to the time step size, using both the FP NM (left) and staging (right) coordinates.}
\label{fig:ECv_dt_obabo_ho_si}
\centering
\end{figure}

\begin{figure}
\includegraphics[width=0.8\textwidth]{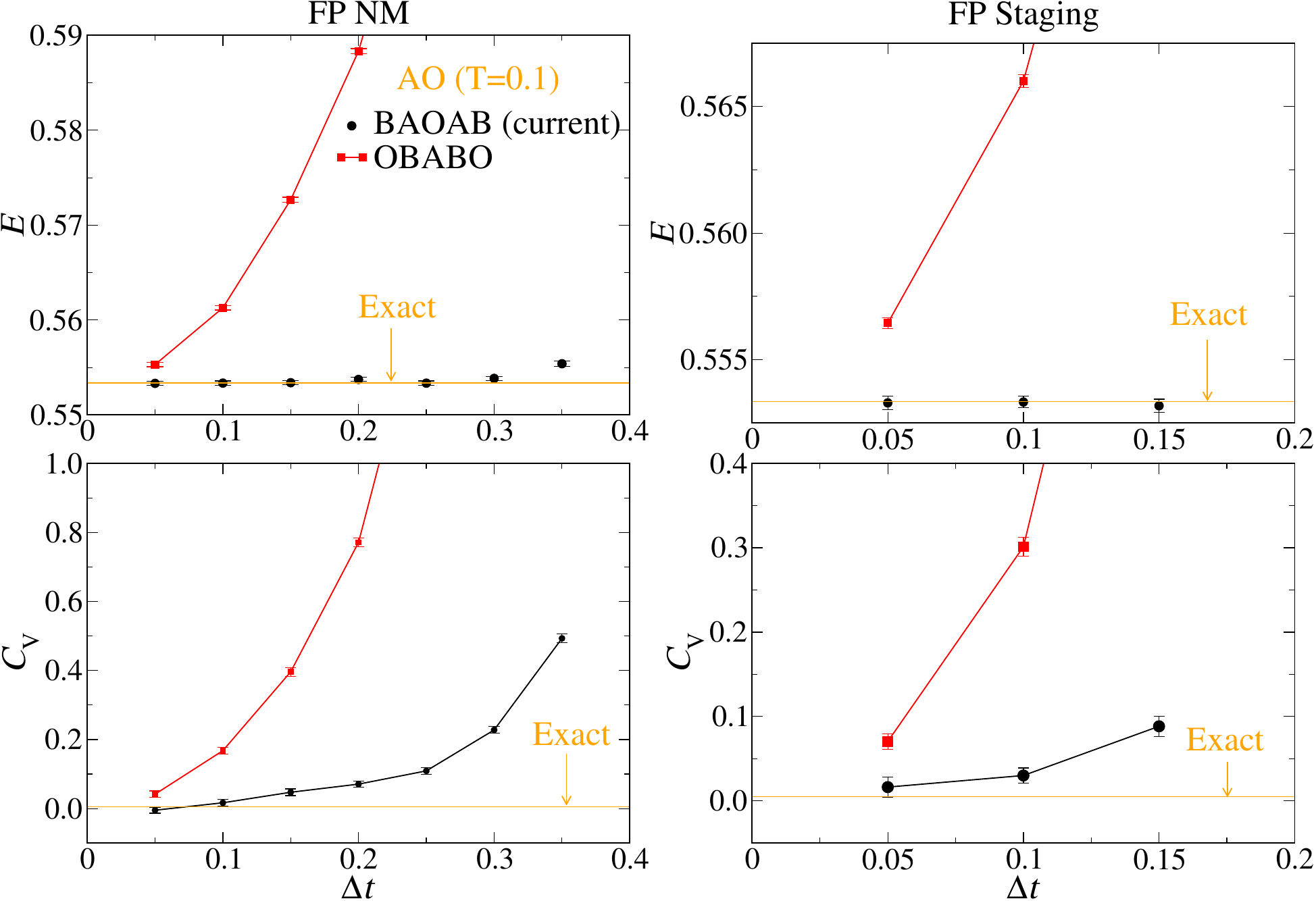}
\caption{Same as Fig.~\ref{fig:ECv_dt_obabo_ho_si}, but for the AO model.}
\label{fig:ECv_dt_obabo_ao_si}
\centering
\end{figure}

Figures~\ref{fig:ECv_dt_obabo_ho_si} and~\ref{fig:ECv_dt_obabo_ao_si} show the effect of the propagator order choices on the convergence of the energy and heat capacity of the HO and AO models, respectively, with respect to time step size. In all cases considered, the BAOAB propagator we adopted in this work shows a remarkable performance over the OBABO alternative. This observation is already established in the literature.\cite{liu2016simple,korol2020dimension} 

\newpage
\bibliography{references}